\documentclass[useAMS,usenatbib,usegraphicx]{mn2e}

\newcommand{\etal}{{et al.\ }}
\newcommand{\be}{\begin{equation}}
\newcommand{\ee}{\end{equation}}
\newcommand{\bea}{\begin{eqnarray}}
\newcommand{\eea}{\end{eqnarray}}
\newcommand\ion[2]{#1$\;${\scshape{#2}}}
\usepackage{amssymb}

\title[Probing the atmosphere of HD 80606b]{Probing potassium in the atmosphere of HD 80606b with tunable filter transit spectrophotometry from the Gran Telescopio Canarias}

\author[K. D. Col\'on \etal]{Knicole D. Col\'on$^{1}$\thanks{E-mail: knicole@astro.ufl.edu}\thanks{NSF Graduate Research Fellow.}, Eric B.\ Ford$^{1}$, Seth Redfield$^{2}$, Jonathan J.\ Fortney$^{3}$, 
\newauthor
Megan Shabram$^{1}$, Hans J. Deeg$^{4,5}$ and Suvrath Mahadevan$^{6,7}$
\\
$^{1}$Department of Astronomy, University of Florida, Gainesville, FL 32611, USA\\
$^{2}$Astronomy Department, Van Vleck Observatory, Wesleyan University, Middletown, CT 06459, USA\\
$^{3}$Department of Astronomy and Astrophysics, University of California, Santa Cruz, CA 95064, USA\\
$^{4}$Instituto de Astrof\'\i sica de Canarias, C. Via Lactea S/N, 38205 La Laguna, Tenerife, Spain\\
$^{5}$Universidad de La Laguna, Dept. de Astrof\'\i sica,  38200 La Laguna, Tenerife, Spain\\
$^{6}$Department of Astronomy \& Astrophysics, Pennsylvania State University, University Park, PA 16802, USA\\
$^{7}$Center for Exoplanets and Habitable Worlds, Pennsylvania State University, University Park, PA 16802, USA}

\begin{document}


\pagerange{\pageref{firstpage}--\pageref{lastpage}} \pubyear{2011}

\maketitle

\label{firstpage}

\begin{abstract} 
We report observations of HD 80606 using the 10.4-m Gran Telescopio Canarias (GTC) and the OSIRIS tunable filter imager.  We acquired very-high-precision, narrow-band photometry in four bandpasses around the \ion{K}{i} absorption feature during the January 2010 transit of HD 80606b and during out-of-transit observations conducted in January and April of 2010.  We obtained differential photometric precisions of $\sim 2.08 \times 10^{-4}$ for the in-transit flux ratio measured at 769.91-nm, which probes the \ion{K}{i} line core.  We find no significant difference in the in-transit flux ratio between observations at 768.76 and 769.91 nm.  Yet, we find a difference of $\sim 8.09 \pm 2.88 \times 10^{-4}$ between these observations and observations at a longer wavelength that probes the \ion{K}{i} wing (777.36 nm).  While the presence of red noise in the transit data has a non-negligible effect on the uncertainties in the flux ratio, the 777.36-769.91 nm colour during transit shows no effects from red noise and also indicates a significant colour change, with a mean value of $\sim 8.99 \pm 0.62 \times 10^{-4}$.  This large change in the colour is equivalent to a $\sim4.2$\% change in the apparent planetary radius with wavelength, which is much larger than the atmospheric scale height.  This implies the observations probed the atmosphere at very low pressures as well as a dramatic change in the pressure at which the slant optical depth reaches unity between $\sim$770 and 777 nm.  We hypothesize that the excess absorption may be due to \ion{K}{i} in a high-speed wind being driven from the exoplanet's exosphere.  We discuss the viability of this and alternative interpretations, including stellar limb darkening, starspots, and effects from Earth's atmosphere.  We strongly encourage follow-up observations of HD 80606b to confirm the signal measured here.  Finally, we discuss the future prospects for exoplanet characterization using tunable filter spectrophotometry.
\end{abstract}


\section{Introduction} 
\label{SecIntro}

Discoveries of extrasolar planets which transit their host star provide valuable opportunities to measure the physical properties of exoplanetary atmospheres.  The physical characteristics of an exoplanetary atmosphere can be probed by transmission spectroscopy observed against the spectrum of the host star.  \citet{seager00}, \citet{brown01} and \citet{hubbard01} developed models that predicted such absorption, particularly from \ion{Na}{i}, \ion{K}{i}, and other alkali metals.  Subsequent refinements of such models have confirmed that in the optical wavelength regime, the strongest lines are expected from the \ion{Na}{i} resonance lines ($\lambda\lambda$ 589.6, 589.0 nm) and the \ion{K}{i} resonance lines ($\lambda\lambda$ 769.9, 766.5 nm) \citep[e.g.,][]{barman07,fortney10}.\footnote[1]{We caution that these lines are most prominent for hot-Jupiter-like planets with a certain range of atmospheric temperatures.  Atmosphere models generated for HD 80606b at the time of transit [based on \citet{fortney10}] do not predict a significant \ion{K}{i} absorption feature, due to the low equilibrium temperature of 500 K.  We refer the reader to \S\ref{secModel} for further discussion.}  In the optical, the cores of the atomic features of \ion{Na}{i} and \ion{K}{i} are relatively narrow.  For this reason, medium to high resolution spectrographs can be used to compare the in-transit stellar spectrum to the out-of-transit stellar spectrum.  The absorption of stellar photons in the exoplanetary atmosphere leads to excess absorption in the in-transit stellar spectrum when compared to the out-of-transit spectrum.   In photometric observations, this leads then to deeper transits and a larger apparent size of the planet at the absorbing wavelengths \citep{brown01}, with variations of order the atmospheric scale height \citep{fortney05}.  Such measurements in strong optical transitions can also constrain the atmospheric metallicity, rainout of condensates, distribution of absorbed stellar flux, and photoionization of atmospheric constituents.

The first detection of absorption due to an exoplanetary atmosphere came from \ion{Na}{i} observations of HD 209458b using the Space Telescope Imaging Spectrograph (STIS) on board the {\it Hubble Space Telescope} ({\it HST}) \citep{charbonneau02}.  Unfortunately, the subsequent failure of the STIS instrument prevented similar observations for more than 5 years.  Thus, attention was directed toward making such observations from the ground \citep[e.g.,][]{moutou01,winn04,narita05}.  The second detection of absorption due to an exoplanetary atmosphere, this time from the ground, was also made of \ion{Na}{i} in observations of HD 189733b using the 9.2-m Hobby-Eberly Telescope (HET) \citep{redfield08}.  Further detections of \ion{Na}{i} in the atmosphere of HD 209458b were made using archival data from the 8.2-m Subaru Telescope \citep{snellen08}, from {\it HST} by \citet{sing08} and from Keck by \citet{langland09}.  The recent repair of STIS and installation of the Cosmic Origins Spectrograph (COS) on board {\it HST} has enabled new optical and ultraviolet (UV) transmission spectrum observations of exoplanetary atmospheres, extended exospheres, and auroral emission \citep[e.g.,][]{linsky10, fossati10, france10}.  

Comparing the surprisingly weak \ion{Na}{i} absorption in HD 209458b \citep{charbonneau02, knutson07} to the ~3 times stronger Na I absorption of HD 189733b \citep{redfield08} suggests that the two planets have different atmospheric structures. Theorists have suggested numerous mechanisms such as adjustments to the metallicity, rainout of condensates, distribution of absorbed stellar flux, or photoionization of sodium \citep{barman07, fortney03}.  In particular, \citet{barman02} suggested that non-LTE Na level populations were the cause of the weak Na feature observed in HD 209458b, and a reanalysis of the \citet{knutson07} data by \citet{sing08, sing08b} suggested that Na condensation or Na photoionization in HD 209458b's atmosphere was the best explanation for matching the data, given the Na line shapes they derived.  It is clear that comparisons of the atmospheric properties of different transiting planets will be critical to understanding the atmospheric properties of exoplanets as a whole.  Although still small, the list of detected atoms and molecules is growing.  In addition to \ion{Na}{i}, several molecules have been detected, primarily in the infrared, with both space-based and ground-based platforms, including CO, CO$_2$, H$_2$O, and CH$_4$ \citep{swain08, swain09,snellen10}.  Other HST observations using the Advanced Camera for Surveys (ACS) did not detect \ion{K}{i} in HD 189733b \citep{pont08}.  If detections of constituents in the extended exosphere are included, then \ion{H}{i}, \ion{C}{ii}, \ion{O}{i}, \ion{Mg}{ii}, and other metals have also been detected \citep{vidalmadjar03, vidalmadjar04, linsky10, fossati10}. 

Each new detection provides not only compositional information, but another window into the physical properties of the exoplanetary atmosphere (e.g., condensation, wind speed, photoionization).  Even though atmosphere models do not predict a significant \ion{K}{i} feature in HD 80606b, it remains of great interest to observationally determine the level of \ion{K}{i} absorption in its atmosphere, since \ion{K}{i} is generally predicted to be the second strongest transmission spectrum signature in the optical wavelength range.  Further, \ion{Na}{i} and \ion{K}{i} probe different layers of the atmosphere.  Measurements of \ion{K}{i} can test the hypothesis that the low abundance of \ion{Na}{i} on HD 209458b may be due to a high altitude layer of clouds or haze. Finding low abundance for both \ion{Na}{i} and \ion{K}{i} would be consistent with either the cloud hypothesis or with the photoionization hypothesis, as both are very easy to ionize.  Finding that only \ion{Na}{i} is significantly depleted would point to alternative models with complex atmospheric chemistry (e.g., incorporation into grains, odd temperature structure, unexpected mixing patterns). Finally, in principle, future observations could probe temporal variability of \ion{Na}{i} and \ion{K}{i} due to high-speed, high-altitude winds and/or differences in the leading and trailing limb \citep{fortney10}.

All of the above atmospheric studies were based on observations using high-resolution spectrographs.  Here, we describe a new technique that utilizes fast, narrow-band, spectrophotometry with the Optical System for Imaging and low Resolution Integrated Spectroscopy (OSIRIS) installed on the 10.4-m Gran Telescopio Canarias (GTC) to probe the composition and other properties of the atmospheres of exoplanets that transit bright stars (see \S\ref{secMethods}).  Fast line spectrophotometry can be much more efficient (e.g., $\sim 34\%$ with GTC/OSIRIS) than typical high resolution spectrographs ($\sim$1-2\%) thanks to the use of a tunable filter (TF) rather than diffraction gratings.  Further, this technique has the potential to be less sensitive to several systematic noise sources, such as seeing variations that cause line-variations in wide spectrograph slits (specifically in non-fiber fed spectrographs), atmospheric variations (since reference stars will be observed simultaneously), and/or flat-fielding errors (since on- and off-line data are obtained at the same detector location).  Thus, spectrophotometry with a TF technique is particularly well-suited for observing a narrow spectral range of atomic absorption features, without suffering from the inefficiencies or potential systematic uncertainties of high-resolution spectrographs.

Here we present results of such observations of the January 2010 transit of HD 80606b using the GTC and the OSIRIS TF imager.  HD 80606b was originally discovered by radial velocity observations \citep{naef01} and was remarkable due to its very high eccentricity ($e=0.93$).  Only several years later did {\em Spitzer} and ground-based observations reveal that the planet passes both behind and in front of its host star \citep{laughlin09, fossey09, garcia09, moutou09}.  Spectroscopic observations revealed that the angular momentum axis of the stellar rotation and that of the orbital planet are misaligned \citep{moutou09, pont09, winn09}.  Given the infrequent transits and long transit duration ($\sim 12$ hours), follow-up observations are quite challenging.  \citet{winn09}, \citet{hidas10} and \citet{shporer10} were able to characterize transits of HD 80606b with longitudinally distributed networks of ground-based observatories, and \citet{hebrard10} observed the January 2010 transit using the {\em Spitzer} spacecraft.  

The {\em Spitzer} observations constrain the thermal properties of the planet's atmosphere \citep{laughlin09, hebrard10}.  To the best of our knowledge, the observations presented here are the first to attempt to detect atmospheric absorption by HD 80606b.  While existing atmosphere models predict that HD 80606b would {\em not} have any significant \ion{K}{i} feature due to its high surface gravity and cold atmosphere at the time of transit (e.g., see \S\ref{secModel}), our observations test this prediction.  Even though models do not predict a \ion{K}{i} feature, exoplanet observations have a track record of unexpected discoveries.  Furthermore, in principle, depending on the atoms/molecules found in the atmosphere, these observations could yield information about how the planet cools, independent of any observations of the thermal phase curve of this system.  In principle, transmission spectroscopy also provides a way to characterize transiting planets in eccentric orbits, which either do not pass behind their host star or which are too cool to detect via occultation when they do pass behind the star.  

Finally, we note that HD 80606 is one of the best systems for making very precise spectrophotometric measurements.  HD 80606 is the brightest of the transiting planet host stars which have a comparably bright reference star very nearby ($\sim$20 arcsec).  Also, the long duration between the 2nd and 3rd points of contact ($\sim$6 hours) of HD 80606b provides time to collect a large amount of in-transit data in a single transit.  Thus, we expect that all else (e.g., observing conditions) being equal, HD 80606b permits the most precise spectrophotometric measurements of any known system (at least with observations of a single transit). 
 
This paper presents extremely precise measurements of the variation in HD 80606b's apparent radius with wavelength near the \ion{K}{i} feature, which in turn can help us test the predictions of atmosphere models.  \S\ref{secMethods} describes our observations and data analysis procedures.  We describe the results of our observations in \S\ref{secResults}.  In \S\ref{secDiscuss} we interpret the results, and we summarize our conclusions and discuss the future prospects for the method in \S\ref{secConclusion} and \S\ref{secFuture}.

\section{Observations}
\label{secMethods}

HD 80606 and its nearby companion (HD 80607) are both bright G5 dwarves of a similar magnitude (V$\sim$9) and colour.  On three nights, we measured the flux of both HD 80606 (target) and HD 80607 (reference) simultaneously.  We cycled through a set of four wavelengths throughout the observations.  On the night of 2010 January 13-14, the planet was in transit for the duration of our observations, and we measure an ``in-transit'' flux ratio of HD 80606 to HD 80607 for each wavelength.  We repeated the observations on 2010 January 15 and 2010 April 4, when the planet was not transiting HD 80606, allowing us to measure the ``out-of-transit'' (OOT) flux ratio of HD 80606 to HD 80607 for each wavelength.  Our results (\S\ref{secResults}) are based on the ratio of in-transit flux ratio (target over reference) to out-of-transit flux ratio (target over reference).  Any changes in the Earth's atmosphere from one night to the next should affect both the target and reference star similarly.  By making differential measurements of the colour during the same transit and at similar atmospheric conditions, this method allows for extremely precise measurements of the transit depth at different wavelengths.  While night-to-night variability in the atmospheric conditions or either of the stars could cause a systematic scaling of the transit depth measurements, the relative wavelength dependence of the apparent planet radius is largely insensitive to either of these potential systematics.  We refer the reader to \S\ref{telluric} and \S\ref{spots} for further discussion.

\subsection{In-Transit and Out-of-Transit Observations}
\label{secObs}

We observed a partial transit of HD 80606b on 2010 January 13-14 and acquired baseline data on 2010 January 15 and 2010 April 4 to establish the OOT flux ratios.  For our observations, we used the TF imaging mode of the OSIRIS instrument installed on the 10.4-m GTC, which is located at the Observatorio del Roque de los Muchachos on the island of La Palma \citep{cepa2000,cepa2003}.  In the TF mode, the user can specify custom bandpasses with a central wavelength of 651-934.5 nm and a FWHM of 1.2-2.0 nm.  The effective wavelength decreases radially outward from the optical centre; because of this effect, we positioned the target and its reference star at the same distance from the optical centre and on the same CCD chip.  The observed wavelengths described below refer to the location of the target (and reference) on the CCD chip.

During the transit observations and baseline observations on 2010 January 15, exposures of the target and its reference star cycled through four different wavelengths (all with a FWHM of 1.2-nm): one on the predicted core of the \ion{K}{i} line (769.75-nm), one to the blue side (768.60-nm) and two redward of the \ion{K}{i} feature (773.50 and 777.20 nm).  As the tunings for the TF are set by the order sorter (OS) filter used, our bluest wavelength is the bluest wavelength we could observe at in the wing of the \ion{K}{i} line and still observe within the same OS filter as the ``on-line'' wavelength (i.e., at the location of the core of the \ion{K}{i} line).  We then chose two wavelengths redward of the \ion{K}{i} line in order to sample more of the structure/wings around the \ion{K}{i} line.  The reddest bandpass was chosen since we expect to see (for a typical hot-Jupiter) a maximum difference between the flux ratio in the on-line bandpass and around that reddest bandpass.  In order to maximize the signal-to-noise ratios in the on-line wavelength and in the reddest off-line wavelength, in each sequence we observed on-line three times, at the reddest off-line wavelength two times and at the other off-line wavelengths one time each.  During the transit, the observing sequence from the GTC was as follows: 769.75, 768.60, 769.75, 773.50, 769.75, 777.20, 777.20 nm (repeat).  

We emphasize that these wavelengths were chosen to be around the location of the \ion{K}{i} feature in HD 80606b's atmosphere.  In order to observe on the \ion{K}{i} feature (which has a rest wavelength of $\sim$769.9 nm) in the frame of the planet, we accounted for the Doppler shifts due to the Earth's motion around the Sun, the system's radial velocity and the planet's non-zero radial velocity during transit [$-59.6$ km s$^{-1}$ based on velocities from \citet{winn09}].  After accounting for these effects, the observed wavelengths in the frame of the planet are redshifted by 0.16 nm to 769.91 nm (on-line) and 768.76, 773.66 and 777.36 nm (off-line).  The observed wavelengths in the frame of the star are essentially the same as observed on Earth due to the small systemic velocity of the HD 80606 planetary system and the Earth's small barycentric velocity on the night of the transit.  For the remainder of the paper, we report the wavelengths as observed in the frame of the planet when discussing results from the transit observations.

A similar sequence as described above was used for the baseline observations taken on 2010 April 4, but the observed wavelengths were corrected for the Doppler shift due to the planet's orbital velocity on that specific date ($\sim$23.9 km s$^{-1}$) in order to match the wavelengths observed during the transit.  Thus, the wavelengths observed on 2010 April 4 (from the GTC) are 770.00 nm (on-line) and 768.86, 773.76 and 777.45 nm (off-line).  

Transit observations of HD 80606b began at 22:28 UT on 2010 January 13 (during ingress) and ended at 7:15 UT on 2010 January 14 (around the beginning of egress and including astronomical twilight), during which the airmass ranged from $\sim$1.08-1.72.  The observing conditions were photometric, with a clear sky and a dark moon.  No data was taken between 5:20 and 5:50 UT on 2010 January 14 due to recalibration of the TF during that time.  The actual seeing varied between 0.7 and 0.9 arcsec during the transit observations, but we used a slight defocus to increase efficiency and reduce the impact of pixel-to-pixel sensitivity variations.  Therefore, the defocussed FWHM of the target varied from $\sim$0.9-2.3 arcsec (7-18 pixels) during the transit.  For the portion of the light curve used in our analysis (see \S\ref{secRed}), the FWHM was much more stable than is indicated by the range given above, with a typical value between 10 and 14 pixels and a mean value of 12 pixels.  Even with an autoguiding system, the target's centroid coordinates shifted by $\sim$9-10 pixels over the course of the night.  We used 1$\times$1 binning and a fast readout mode (500 kHz) to readout a single window of 300 $\times$ 600 pixels (located on one CCD chip) in order to reduce the dead time between exposures.  This window is equivalent to a field-of-view of $\sim$ 38$\times$76 arcsec, so the only stars in our field were HD 80606 and a single reference star, HD 80607.  Each individual observation was followed by an average dead time of less than 4-s for readout and to switch between TF tunings.  We used 10-s exposures, resulting in an overall cadence of about 14-s for each observation.  Due to the short exposure time used, the sky background level was low enough that we did not need to discard any images taken during astronomical twilight.

Baseline observations were taken from 5:50 to 7:10 UT (i.e., also through the beginning of astronomical twilight) on 2010 January 15, but the data was highly scattered, so we do not include it in our primary analysis.\footnote[2]{See \S\ref{secCompare} for further discussion of this data set.}  Additional baseline observations took place on 2010 April 4 from 21:30 (including the end of astronomical twilight) to 0:00 UT.  The observing conditions were photometric and taken during grey time, using the same set up as the in-transit observations described above.  During the observations, the airmass ranged from $\sim$1.08-1.20, and the actual seeing varied between 1.4 and 1.6 arcsec (11-12.5 pixels), so the telescope was not intentionally defocussed.  The target's centroid coordinates shifted by $\sim$5-8 pixels during the observations.  The exposure time was changed from the initial exposure time of 10-s to 8-s and then again to 11-s to counteract variations in the seeing as well as increasing airmass while avoiding saturation and maintaing a high number of counts.  In our analysis, we discard the 10-s data because a majority of the images were saturated.  We tested using the OOT flux ratios from the 8 and 11 s data individually in our analysis and found that they produced very similar results.  Thus, we combine the 8 and 11 s data to establish the final OOT flux ratios (see \S\ref{secRed}) and to achieve the longest usable baseline possible.

\subsection{Data Reduction and Analysis}
\label{secRed}

Observations taken with OSIRIS prior to mid-March 2010 suffered from a higher than expected level of dark current despite the short exposure times used.  Therefore, we used standard IRAF procedures for bias and dark subtraction and flat field correction for the January 2010 transit observations of HD 80606.  We note that the flat fields for these observations did not produce the pattern of having the total number of counts in the dome flat fields decreasing with time as seen by \citet{colon10}, so we use almost all (65 out of 75) dome flats for each observed wavelength in our analysis (the 10 dome flats not included in the analysis were overexposed).  A new dewar fixed the problems with the dark current before the April 2010 observations took place, so for the baseline data we performed standard bias subtraction and flat field correction (combining all 133 flats taken for each observed wavelength) and did not need to subtract dark frames.  

\begin{figure}
\includegraphics[width=84mm]{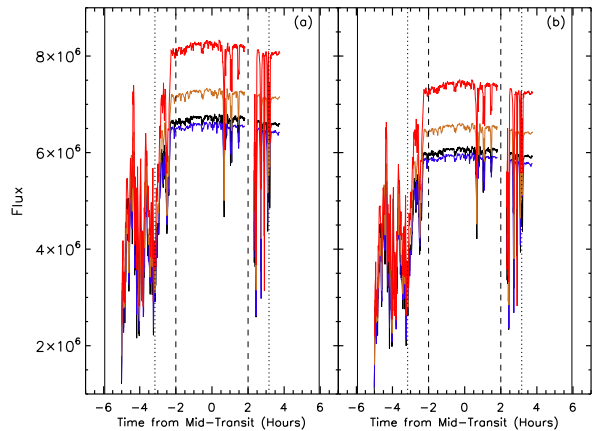}
\caption{Absolute fluxes of HD 80606 (a) and HD 80607 (b) as measured on 2010 January 13-14.  The different light curves represent the fluxes as measured nearly simultaneously in the different bandpasses, with the black, blue, brown and red light curves representing the 769.91, 768.76, 773.66 and 777.36 nm data.  These data have not been corrected for airmass or decorrelated in any way.  Note the break in the data around two hours after mid-transit due to recalibration of the tunable filter.  The vertical solid lines indicate the expected beginning and end of the transit, and the vertical dotted lines indicate the end of ingress and the beginning of egress [based on durations estimated by \citet{hebrard10} and the transit ephemeris from \citet{shporer10}].  The vertical dashed lines indicate the $\sim$4 hour interval around mid-transit that our analysis focused on (see text for further details).}
\label{figjanflux}
\end{figure}

\begin{figure}
\includegraphics[width=84mm]{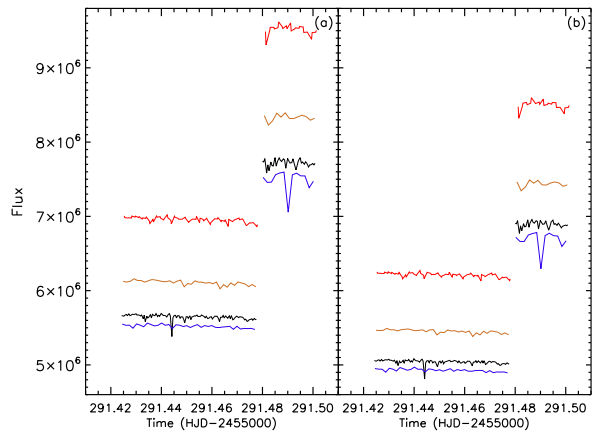}
\caption{Similar to Figure \ref{figjanflux}, but for the out-of-transit data taken the night of 2010 April 4.  Note that the discontinuity in the fluxes around 2455291.48 is due to a change in the exposure time (from 8-seconds to 11-seconds).}
\label{figaprflux}
\end{figure}

Because of the very small readout window used for our observations, our images do not contain the sky (OH) emission rings that occur due to the TF's small bandpass and position-dependent wavelength.  Therefore, we performed simple aperture photometry on the target and reference star using the standard IDL routine APER\footnote[3]{http://idlastro.gsfc.nasa.gov/} for a range of aperture radii.  We measured the root mean square (RMS) scatter of the flux ratio (equal to the target flux divided by the reference flux) at the bottom of the transit (for the January 2010 data) and for the individual 8 and 11 s data taken OOT (in April 2010) in each bandpass.  We considered the results for each bandpass and adopted an aperture radius of 28 pixels (3.6 arcsec) for the in-transit data and 32 pixels (4.1 arcsec) for the OOT data, as these were the aperture radii that typically yielded the lowest RMS scatter.  The radii of the sky annulus used for the reduction of both data sets was 68 to 74 pixels in order to completely avoid any flux from the target or reference star.  We have included the results of our aperture photometry in Tables \ref{tabjanin} and \ref{tabaprout} and illustrate the results in Figures \ref{figjanflux} and \ref{figaprflux}.  As illustrated, the flux in each bandpass displayed large variations during parts of the observations (particularly during parts of the transit), and we take this into consideration in our analysis (see \S\ref{secearth}).  

\begin{table}
  \caption{Absolute Transit Photometry from 2010 January 13 \label{tabjanin} }
  \begin{tabular}{@{}cccc@{}}
  \hline
$\lambda$ (nm) & HJD & $F_{target}$ & $F_{ref}$   \\
\hline
768.76 & 2455210.4428 & 2789803 & 2501667 \\
\multicolumn{4}{c}{\dots} \\
\multicolumn{4}{c}{} \\
769.91 & 2455210.4414 & 1207952 & 1081769  \\
\multicolumn{4}{c}{\dots} \\
\multicolumn{4}{c}{} \\
773.66 & 2455210.4419 & 1789988 & 1603033  \\
\multicolumn{4}{c}{\dots} \\
\multicolumn{4}{c}{} \\
777.36 & 2455210.4423 & 2441876 & 2188980  \\
\multicolumn{4}{c}{\dots} \\
\hline
\end{tabular}

\medskip
The wavelengths included in the table are the observed wavelengths in the frame of the planet (see text for additional details).  The time stamps included here are for the times at mid-exposure.  $F_{target}$ and $F_{ref}$ are the absolute flux measurements of HD 80606 and HD 80607.  The full table is included online, while a portion is shown here so the reader can see the formatting of the table.

\end{table}

 \begin{table}
  \caption{Absolute OOT Photometry from 2010 April 4 \label{tabaprout} }
  \begin{tabular}{@{}ccccc@{}}
  \hline
$\lambda$ (nm) & $t_{exp}$ (s) & HJD & $F_{target}$ & $F_{ref}$ \\
\hline
768.86 & 8 & 2455291.4245 & 5550569 & 4951634  \\
\multicolumn{5}{c}{\dots} \\
\multicolumn{5}{c}{} \\
770.00 & 8 & 2455291.4243 & 5665554 & 5060730 \\
\multicolumn{5}{c}{\dots} \\
\multicolumn{5}{c}{} \\
773.76 & 8 & 2455291.4249 & 6125160 & 5466145 \\
\multicolumn{5}{c}{\dots} \\
\multicolumn{5}{c}{} \\
777.45 & 8 & 2455291.4253 & 7006625 & 6252784 \\
\multicolumn{5}{c}{\dots} \\
\hline
\end{tabular}

\medskip
Columns are similar to Table \ref{tabjanin}, except the wavelengths included in the table are the wavelengths as observed from the GTC (see text for additional details).  The second column contains the exposure time for the observations, as observations based on two different exposure times were included in our analysis.  The full table is available online, and a portion is shown here so the reader can see the formatting of the table.

\end{table}

\begin{figure}
\includegraphics[width=84mm]{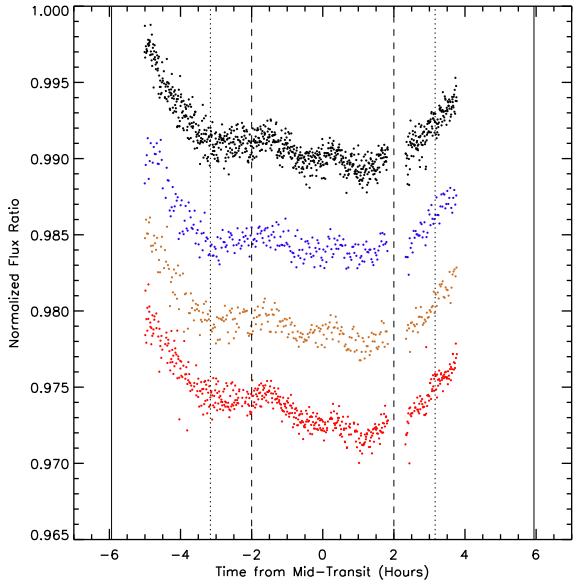}
\caption{Transit light curves as observed nearly simultaneously in different bandpasses on 2010 January 13-14.  The on-line light curve (769.91-nm) is shown in black, and the off-line light curves (768.76, 773.66 and 777.36 nm) are shown in blue, brown and red.  The flux ratio for each bandpass has been normalized to the weighted mean OOT flux ratio estimated from the baseline data acquired in April 2010, but the data has not been corrected for airmass or decorrelated in any way.  The off-line light curves have been arbitrarily offset by 0.006, 0.012 and 0.018, and error bars are not shown for clarity.  The vertical solid, dotted and dashed lines are the same as in Figure \ref{figjanflux}.}
\label{figlcfull}
\end{figure}

\begin{figure}
\includegraphics[width=84mm]{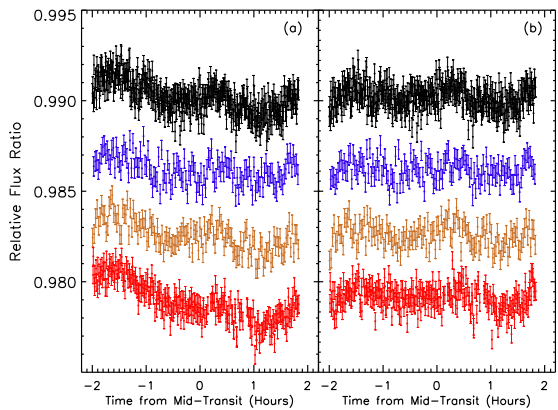}
\caption{Relative in-transit flux ratio normalized to the relative OOT flux ratio as measured on 2010 April 4.  The relative flux before (a) and after (b) EPD was applied is shown.  The different colors represent the flux ratios as measured in the different bandpasses, with the colors the same as in Figure \ref{figlcfull}.  Note that EPD was only applied to the $\sim$4 hours centered around mid-transit (i.e. the bottom of the transit light curve).  The data shown has not been binned, but the different light curves have been offset arbitrarily for clarity.}
\label{figinepds}
\end{figure}

\begin{figure}
\includegraphics[width=84mm]{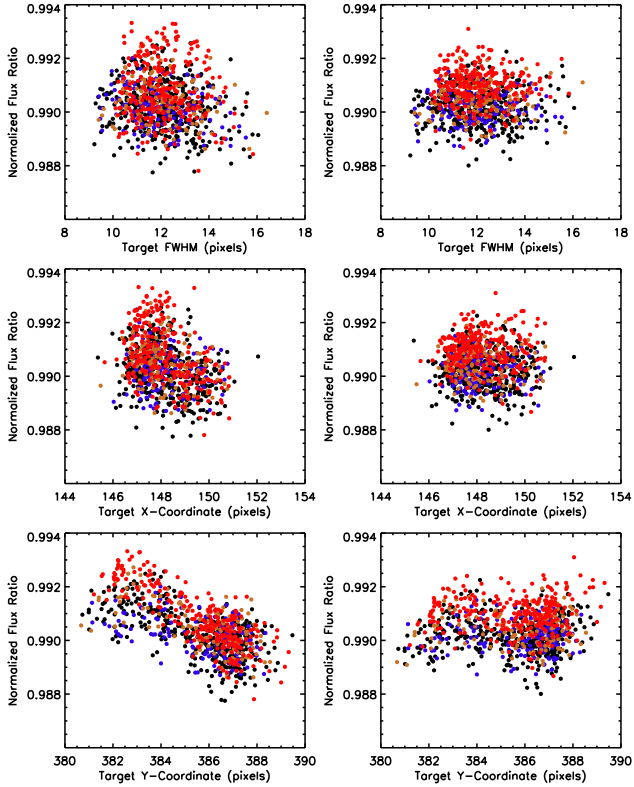}
\caption{Correlations between the normalized in-transit flux ratio and the target FWHM and $x$ and $y$ centroid coordinates, before (lefthand column) and after (righthand column) EPD has been applied.  All four bandpasses are shown in each panel, with the colours the same as in Figure \ref{figlcfull}.  Similar results were obtained when decorrelating the data against the reference parameters but are not shown here.}
\label{figlccorrs}
\end{figure}

\begin{figure}
\includegraphics[width=84mm]{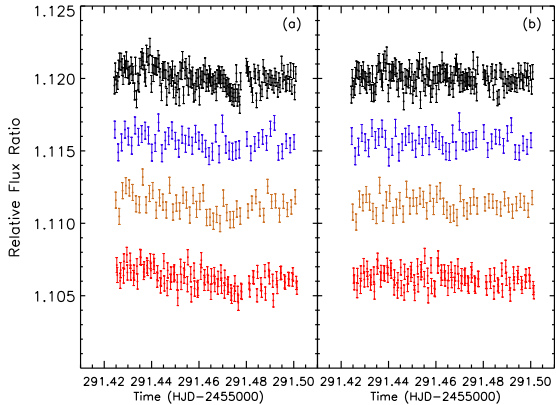}
\caption{Relative OOT flux ratio as measured on 2010 April 4.  The relative flux before (a) and after (b) EPD was applied is shown.  The different colors represent the flux ratios as measured in the different bandpasses, with the colors the same as in Figure \ref{figlcfull}.  Note the small break in the data around 2455291.48 where the exposure time was changed.  The data has not been binned, but the different light curves have been offset arbitrarily for clarity.}
\label{figoutepds}
\end{figure}

We present the raw in-transit light curves in Figure \ref{figlcfull}, which were computed by dividing the flux in the target aperture by the flux in the reference star aperture and then normalizing by the weighted mean OOT flux ratio for each bandpass (see \S\ref{secResults} for details on the computation of the mean flux ratios).  In an attempt to reduce systematic trends seen in our transit light curves, we applied the external parameter decorrelation (EPD) technique \citep[see, e.g.,][]{bakos07,bakos10} to each transit and baseline light curve.  Note that for the transit light curve, we only applied EPD to the $\sim$4 hours centered around mid-transit, or 3:36 UT on 2010 January 14, as estimated by \citet{shporer10}.\footnote[4]{This ephemeris is in between that given by \citet{winn09} and \citet{hebrard10}.  The choice of ephemeris used does not significantly affect our results.}  Specifically, we decorrelated each individual light curve against the following parameters: the centroid coordinates of both the target and reference, the sharpness of the target and reference profiles [equivalent to (2.35/FWHM)$^2$] and the airmass.  As illustrated in Figure \ref{figinepds}, EPD removed most of the correlations in the in-transit data.  For reference, we show the correlations between the in-transit data and the target's FWHM and centroid coordinates both before and after EPD has been applied in Figure \ref{figlccorrs}.  For the baseline data, we performed EPD for the 8 and 11 s data series separately, but we then combined the two data sets to compute the weighted mean flux ratio and its uncertainty for each bandpass as described in \S\ref{secResults}.  The results of the decorrelation for the OOT data are illustrated in Figure \ref{figoutepds}.  As a result of applying EPD, the RMS scatter in each of the bandpasses improved by as much as $\sim$25\%, but decorrelating the light curves against the above parameters did not completely remove the systematics that are seen in our data.  In a further attempt to remove systematics, we also tried a quadratic decorrelation against the sharpness of the target and reference profiles, as that was the only parameter that showed a possible residual systematic pattern after EPD was applied.  However, the quadratic decorrelation did not reduce systematics in our light curves any further.  We discuss other potential sources of systematics in detail in \S\ref{systematics}.  

\begin{figure}
\includegraphics[width=84mm]{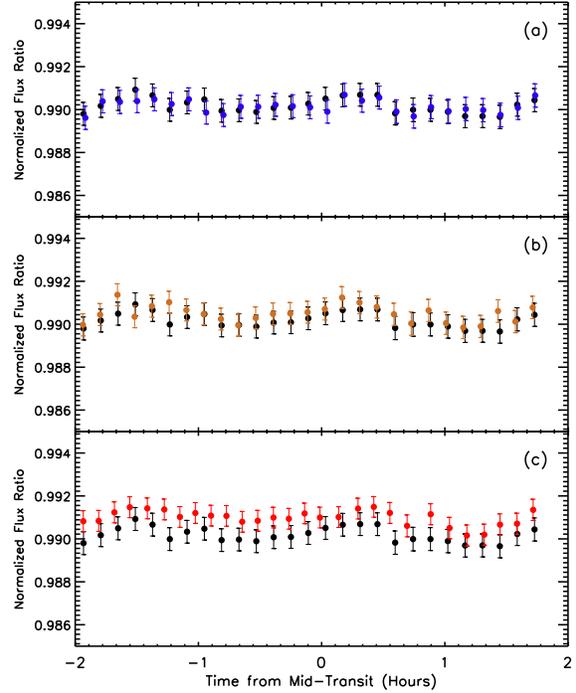}
\caption{Corrected light curves for observations of the bottom of the transit as observed nearly simultaneously in different bandpasses on 2010 January 13-14.  In each panel, the black points illustrate the measurements taken in the on-line (769.91-nm) bandpass.  We also show measurements taken in each of the off-line bandpasses (768.76, 773.66, 777.36 nm) in each of the respective panels (a, b, c) for comparison to the on-line flux ratios.  The data shown here has been decorrelated.  The colours and normalizations are the same as in Figure \ref{figlcfull}, but no offsets have been applied.  Here, we have binned the data and error bars simply for clarity.}
\label{figlcbottom}
\end{figure}

Because our goal is to compare the depths of the transit in each bandpass, the rest of our analysis focuses on the data from the bottom of the transit as presented in Figure \ref{figlcfull} and highlighted in Figure \ref{figlcbottom} i.e., the $\sim$4 hours centered around mid-transit.  Note that the light curves shown in Figure \ref{figlcbottom} have been corrected using EPD.  We also discarded points that had a flux ratio greater than 3$\sigma$ from the mean of the bottom of the transit light curve.  This resulted in discarding four points from the reddest light curve (777.36-nm).  We also discarded several exposures from each wavelength that were unusable due to saturation.  The different panels in Figure \ref{figlcbottom} illustrate the deviation between the magnitude of the on-line flux ratios and each of the off-line flux ratios, which will be discussed in detail in \S\ref{secResults} and \S\ref{secDiscuss}.  

We estimated the uncertainties in the flux ratios by computing the quadrature sum of the photon noise for HD 80606 and HD 80607, the uncertainty in the sum of the sky background (and dark current, for the in-transit observations) and the scintillation noise for the two stars.  We assume Poisson statistics to compute the uncertainty in the sky background, and the noise due to scintillation was estimated from the relation given by \citet{dravins98}, based on \citet{young67}.  We caution that this empirical relation might overestimate scintillation for large telescopes located at excellent sites such as La Palma.  Regardless, the relation demonstrates that scintillation is still a small contribution to the total error budget for these observations.  The flat field noise is also negligible compared to the photon noise, so we do not include it in our determination of the measurement uncertainties.  Based on the relation given by \citet{howell06}, which computes the standard deviation of a single measurement in magnitudes and includes a correction term between the error in flux units and the error in magnitudes, we find the median total uncertainties in the flux ratio for each exposure to be 0.538, 0.532, 0.514 and 0.486 mmag at 768.76, 769.91, 773.66 and 777.36 nm (over the bottom of the transit).  The RMS of the transit light curve is comparable, but slightly larger, with values of 0.585, 0.667, 0.631 and 0.662 mmag for those wavelengths.  The median total uncertainties for the OOT observations are calculated in a similar way, but the uncertainties for the 8 and 11 s data sets were scaled by the flux ratios for each respective set in order to compute a weighted uncertainty.  Thus, the median total (weighted) uncertainties in the flux ratio are 0.657, 0.650, 0.627 and 0.592 mmag while the estimated RMS is quite comparable, with values of 0.562, 0.605, 0.554 and 0.586 mmag for 768.73, 769.87, 773.63 and 777.32 nm.    
 
The complete photometric time series for each bandpass of the in-transit data (uncorrected and unnormalized) is reported in Table \ref{tabinjan1}, while the photometric time series (both before and after EPD was applied) for the transit bottom and the April observations are reported in Tables \ref{tabinjan2} and \ref{tabootapr}.  The weighted mean flux ratios for both the in-transit and OOT data (see \S\ref{secResults} for more details) are given in Table \ref{tabratio}, along with their uncertainties. 

\begin{table}
  \caption{Relative Transit Photometry   \label{tabinjan1} }
  \begin{tabular}{@{}cccc@{}}
  \hline
$\lambda$ (nm) & HJD & $F_{ratio}$ & Uncertainty \\
\hline
768.76 & 2455210.4428 & 1.11518 & 0.00086 \\
\multicolumn{4}{c}{\dots} \\
\multicolumn{4}{c}{} \\
769.91 & 2455210.4414 & 1.11665 & 0.00124 \\
\multicolumn{4}{c}{\dots} \\
\multicolumn{4}{c}{} \\
773.66 & 2455210.4419 & 1.11663 & 0.00104 \\
\multicolumn{4}{c}{\dots} \\
\multicolumn{4}{c}{} \\
777.36 & 2455210.4423 & 1.11553 & 0.00091 \\
\multicolumn{4}{c}{\dots} \\
\hline
\end{tabular}

\medskip
The wavelengths included in the table are the observed wavelengths in the frame of the planet (see text for additional details).  The time stamps included here are for the times at mid-exposure.  $F_{ratio}$ represents the relative flux ratio between the target and reference star (i.e., $F_{target}/F_{ref}$).  The full table is available online, and a portion is shown here so the reader can see the formatting of the table.

\end{table}

\begin{table}
  \caption{Normalized Photometry from around Mid-Transit  \label{tabinjan2} }
  \begin{tabular}{@{}ccccc@{}}
  \hline
$\lambda$ (nm) & HJD & $F_{ratio}$ & $F_{ratio}$ & Uncertainty \\
\multicolumn{1}{c}{} & \multicolumn{1}{c}{} & \multicolumn{1}{c}{$(raw)$} & \multicolumn{1}{c}{$(corrected)$} & \multicolumn{1}{c}{} \\
\hline
768.76 & 2455210.5680 & 0.99033 & 0.98962 & 0.00054 \\
\multicolumn{5}{c}{\dots} \\
\multicolumn{5}{c}{} \\
769.91 & 2455210.5670 & 0.99041 & 0.98912 & 0.00054 \\
\multicolumn{5}{c}{\dots} \\
\multicolumn{5}{c}{} \\
773.66 & 2455210.5671 & 0.99036 & 0.98909 & 0.00052 \\
\multicolumn{5}{c}{\dots} \\
\multicolumn{5}{c}{} \\
777.36 & 2455210.5675 & 0.99228 & 0.99070 & 0.00049 \\
\multicolumn{5}{c}{\dots} \\
\hline
\end{tabular}

\medskip
The wavelengths included in the table are the observed wavelengths in the frame of the planet (see text for additional details).  The time stamps included here are for the times at mid-exposure.  The flux ratios are presented both before (raw) and after (corrected) EPD was applied.  The flux ratios have also been normalized to the weighted mean OOT flux ratio (see Table \ref{tabratio} and text for more details).  The full table is available online, and a portion is shown here so the reader can see the formatting of the table.

\end{table}

\begin{table*}
\centering
\begin{minipage}{140mm}
  \caption{Relative OOT Photometry from 2010 April 4 \label{tabootapr} }
  \begin{tabular}{@{}cccccc@{}}
  \hline
$\lambda$ (nm) & $t_{exp}$ (s) & HJD & $F_{ratio}$ & $F_{ratio}$ & Uncertainty \\
\multicolumn{1}{c}{} & \multicolumn{1}{c}{} & \multicolumn{1}{c}{} & \multicolumn{1}{c}{$(raw)$} & \multicolumn{1}{c}{$(corrected)$} & \multicolumn{1}{c}{} \\
\hline
768.86 & 8 & 2455291.4245 & 1.12096 & 1.12050 & 0.00058 \\
\multicolumn{6}{c}{\dots} \\
\multicolumn{6}{c}{} \\
770.00 & 8 & 2455291.4243 & 1.11951 & 1.11888 & 0.00058 \\
\multicolumn{6}{c}{\dots} \\
\multicolumn{6}{c}{} \\
773.76 & 8 & 2455291.4249 & 1.12056 & 1.11973 & 0.00056 \\
\multicolumn{6}{c}{\dots} \\
\multicolumn{6}{c}{} \\
777.45 & 8 & 2455291.4253 & 1.12056 & 1.11983 & 0.00053 \\
\multicolumn{6}{c}{\dots} \\
\hline
\end{tabular}

\medskip
The wavelengths included in the table are the wavelengths as observed from the GTC (see text for additional details).  The time stamps included here are for the times at mid-exposure.  The flux ratios are presented both before (raw) and after (corrected) EPD was applied.  The full table is available online, and a portion is shown here so the reader can see the formatting of the table. 

\end{minipage}
\end{table*}


\section{Results}
\label{secResults}

As illustrated in Figure \ref{figlcbottom}, we can see by eye a hint of a deviation between the in-transit flux ratios observed at the on-line wavelength and the red off-line wavelengths, but no clear deviation is seen when compared to the bluest off-line wavelength.  Despite evidence of time-correlated systematics in our data, we emphasize that the error bars shown in Figure \ref{figlcbottom} are binned error bars, which illustrate that our measurement uncertainties are larger than any residual systematics present in the light curves and that the deviations in the flux ratios between the different bandpasses are real.  We refer the reader to our discussion of possible systematic sources in \S\ref{systematics}.  

\begin{figure}
\includegraphics[width=84mm]{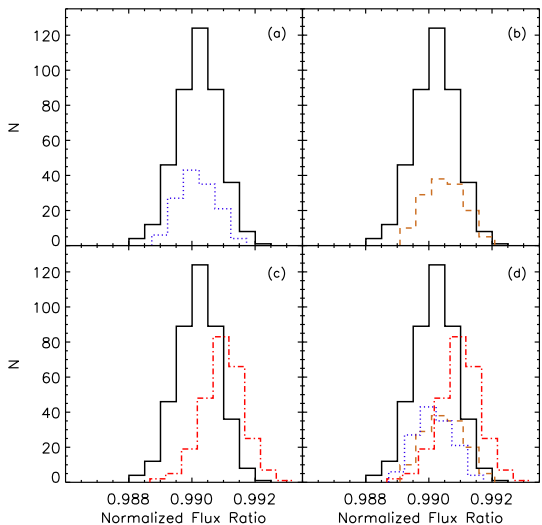} 
\caption{Histograms of normalized flux ratios from the bottom of the transit light curve as shown in Figure \ref{figinepds}$b$.  The histograms were generated using a bin size of 0.5 mmag.  Each panel compares the on-line flux ratios with the off-line flux ratios.  In each panel, the black (solid) histograms represent the 769.91-nm (on-line) light curve.  The blue (dotted), brown (dashed) and red (dot-dashed) histograms are for the 768.76, 773.66 and 777.36 nm light curves and are shown in panels (a), (b) and (c), respectively.  Panel (d) shows the histograms for all four wavelengths for further comparison.}
\label{fighisto}
\end{figure}

In Figure \ref{fighisto}, we plot histograms of the (unbinned) flux ratios at the bottom of each of the transit light curves, where the flux ratios have been normalized against the mean OOT flux ratio for each respective wavelength.  These histograms further illustrate that the flux ratios for the on-line and bluest off-line light curves are comparable, but the red off-line flux ratios (particularly for the reddest light curve) clearly lie at slightly higher values compared to the on-line flux ratios, indicating a smaller apparent planetary radius at those wavelengths.  

Ideally, when one has access to either a complete or partial transit light curve and baseline data acquired immediately before or after the transit event, one can fit a model to the data and estimate the transit depth from the model results.  Due to the very long duration of HD 80606b's transit, we were not able to acquire baseline data on the night of the transit, thereby making this type of analysis impractical. However, thanks to several recent campaigns to observe a complete transit of HD 80606b and establish accurate orbital and physical parameters for this system via light curve modeling \citep{winn09, hebrard10, hidas10, shporer10}, we do not need to fit a model to our partial light curve to achieve the goals of this paper.  Instead, we consider only the middle $\sim$4 hours of the transit light curve in our analysis (compared to the full duration of the bottom of the transit, which is $\sim$6 hours), thereby minimizing systematic effects of stellar limb darkening (LD) as the strongest LD occurs during ingress, egress, and right after/before ingress/egress.  Further, since we do not know the LD model for this star to the precision of our observations, adding such a model would not be useful for this study.  Thus, we assume that LD is the same over all our bandpasses and that the transit ephemeris, impact parameter and transit duration do not vary with wavelength.  The only parameter of which we assume changes with wavelength is the apparent planet radius ($R_{p}$).  

To investigate how the apparent planet radius changes with wavelength, we simply compute the weighted mean in-transit flux ratio [$< \delta F/F >$, which is proportional to the planet-star radius ratio, $(R_p / R_{\star})^2$] and its uncertainty for each wavelength.  Specifically, we compute the weighted mean as
\be
< \delta F/F >=\frac{\displaystyle\sum^{n}_{i=1}w_iF_i}{\displaystyle\sum^{n}_{i=1}w_i}
\ee
where the weights, $w_i$, are equal to 1/$\displaystyle(\beta\sigma_{i})^{2}$.  Here, $\sigma_i$ is the estimated photometric uncertainty weighted by some wavelength-specific factor ($\beta$) in order to account for the presence of any red noise in each individual bandpass.  

\begin{figure}
\includegraphics[width=84mm]{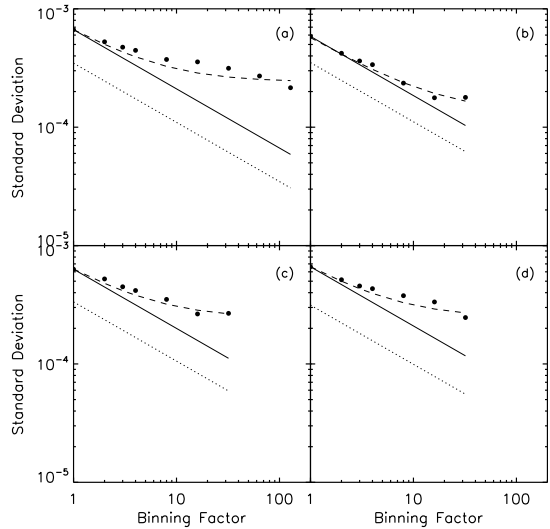} 
\caption{Standard deviation of the time-binned flux ratio measurements from the bottom of the transit (e.g. as shown in Figure \ref{figinepds}$b$) as a function of the number of data points per bin ($N$).  Panels (a), (b), (c) and (d) show the standard deviations for the binned 769.91, 768.76, 773.66 and 777.36 nm light curves.  The amount of binning that could be performed varies for each light curve since the different wavelengths were observed a different number of times in a given observing sequence (see \S\ref{secObs} for details).  The solid line in each panel represents the trend expected for pure white Gaussian noise ($\sim N^{-1/2}$), normalized to the unbinned standard devation measured in our data.  The dotted lines represent the trend for Gaussian noise when normalized to the theoretical noise for our observations.  The dashed curves are models fit to the standard deviation that include both white and red noise.  The effect of red noise is obvious in all bandpasses.}   
\label{figrmsin}
\end{figure}

\begin{figure}
\includegraphics[width=84mm]{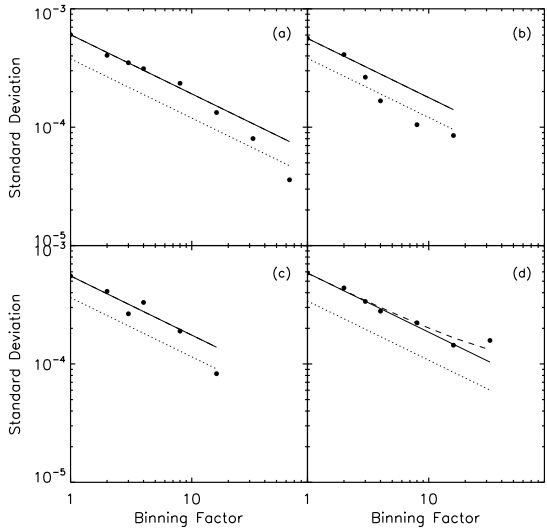} 
\caption{Standard deviation of the time-binned OOT flux ratio measurements from April 2010 (e.g. as shown in Figure \ref{figoutepds}$b$) as a function of the number of data points per bin ($N$).  Panels (a), (b), (c) and (d) show the standard deviations for the binned 770.00, 768.86, 773.76 and 777.45 nm light curves.  The solid line in each panel represents the trend expected for pure white Gaussian noise ($\sim N^{-1/2}$).  The dotted lines represent the trend for Gaussian noise when normalized to the theoretical noise for our observations.  The dashed curves are models fit to the standard deviation that include both white and red noise.  Compared to the in-transit observations, red noise has a very minimal effect here.  Deviations below the curve are likely due to small number statistics.  These results demonstrate that narrow-band ground-based observations can provide very high-precision differential photometry.  For a given bandpass, the combined precision exceeds that of Spitzer \citep{hebrard10} or HST observations \citep{pont08}.  To the best of our knowledge, these represent the highest precision photometry for a 1.2-nm bandpass for ground or space observations.}   
\label{figrmsout}
\end{figure}

To illustrate the effect of red noise on our measurements and the need for a re-weighting factor, the standard deviations ($\sigma_{N}$) of the in-transit and OOT time-binned flux ratios are shown in Figures \ref{figrmsin} and \ref{figrmsout} as a function of binning factor ($N$) for each bandpass.  The theoretical trend expected for white Gaussian noise ($\sim N^{-1/2}$) is plotted as a solid curve, and we can see that for the in-transit data, the RMS deviates from the theoretical curve at large binning factors, indicating that red noise is present in most bandpasses (being the least significant in the bluest bandpass).  However, for the OOT data, our photometry appears to be generally consistent with the photon limit (although the bluest light curve suffers from small number statistics).

Following methods used by, e.g.,  \citet{pont2006} and \citet{winn2007}, we calculated explicit estimates for both the white ($\sigma_w$) and red ($\sigma_r$) noise in each bandpass by solving the following system of equations:
\be
\sigma^{2}_{1}=\sigma^{2}_{w}+\sigma^{2}_{r}
\ee
\be
\sigma^{2}_{N}=\frac{\sigma^{2}_{w}}{N}+\sigma^{2}_{r}.
\ee
The re-weighting factor, $\beta$, is then computed as $\sigma_{r}/(\sigma_{w}/\sqrt{N})$.  Based on our fits to the red and white noise, we computed a re-weighting factor for each bandpass and applied it as stated above.  We imposed a minimum value for $\beta$ of 1, particularly for cases where red noise was negligible.

\begin{figure}
\includegraphics[width=84mm]{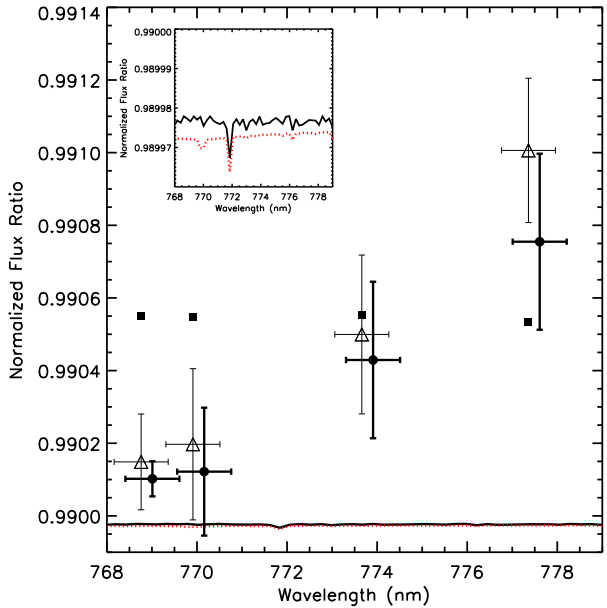} 
\caption{Normalized weighted mean in-transit flux ratio versus observed wavelength (in the frame of the planet).  The open triangles represent the flux ratios as computed for each light curve described in \S\ref{secMethods} and \ref{secResults}.  The solid circles represent the flux ratios computed after excluding outlying absolute flux values for each star from the analysis (see \S\ref{secearth}).  Note that the solid circles have been offset by 0.25-nm for clarity.  The vertical error bars include a factor to account for the effects of red noise in both the in- and out-of-transit data.  The ``error bars'' in the horizontal direction indicate the FWHM of each bandpass.  The solid squares represent the mean in-transit flux ratios estimated from limb-darkened transit light curve models for HD 80606b.  The lines show the predictions of planetary atmosphere models (see \S\ref{secModel} for more details).  The inset figure shows the atmosphere models on a small vertical scale.  While LD or night-to-night variability (of Earth's atmosphere or either star) could affect the overall normalization, the observed change in the flux ratio with wavelength is robust.}  
\label{figspectrum}
\end{figure}

The uncertainties for the OOT flux ratio are also weighted by the flux ratio, $F_{i}$, since two different exposure times were used during the OOT observations.  Finally, the uncertainty on the weighted mean is computed as
\be
\sigma_{< \delta F/F >}=\sqrt{\frac{1}{\displaystyle\sum^{n}_{i=1}w_i}}.
\ee
We include the uncertainty on the weighted mean OOT flux ratio in our calculation of the mean normalized in-transit flux ratio and its uncertainty.  The resulting spectrum of HD 80606b (the normalized weighted mean in-transit flux ratios as a function of wavelength) is shown in Figure \ref{figspectrum}, and it clearly illustrates a difference between the flux ratios for the bluest bandpasses and those for the reddest bandpasses.  While we find no significant difference between the flux ratios measured at 768.76 and 769.91 nm, we measure differences of $3.02 \pm 3.02 \times 10^{-4}$ and $8.09 \pm 2.88 \times 10^{-4}$ between observations at 769.91 and 773.66 and 777.36 nm.

We list the weighted mean in-transit flux ratios (normalized by the weighted mean OOT flux ratios) as well as the weighted mean OOT flux ratios and their uncertainties in Table \ref{tabratio}.  In this table, we also include our fits to the white and red noise, as well as our estimates for $\beta$.  When calculating the normalized in-transit flux ratio and its uncertainty, we also included the re-weighted uncertainty for the mean OOT flux ratio in our calculation.  The error bars for the flux ratios given in Table \ref{tabratio} and shown in Figure \ref{figspectrum} also take red noise into account.  

\begin{table*}
 \centering
 \begin{minipage}{140mm}
  \caption{Time-Averaged Flux Ratios and Noise Estimates \label{tabratio}}
  \begin{tabular}{@{}cccccccc@{}}
  \hline
$\lambda_{E}$ (nm) & $\lambda_{P}$ (nm) & $\lambda_{S}$ (nm) & $< \delta F/F >$ & $\sigma_{< \delta F/F >}$ & $\sigma_{w}$ & $\sigma_{r}$ & $\beta$ \\
\hline
\hline
\multicolumn{8}{c}{In-Transit} \\
\hline
768.60 & 768.76 & 768.60 & 0.9901486 & 1.32 $\times 10^{-4}$ & 5.55 $\times 10^{-4}$ & 1.34 $\times 10^{-4}$ & 2.81 \\
769.75 & 769.91 & 769.75 & 0.9901971 & 2.08 $\times 10^{-4}$ & 6.21 $\times 10^{-4}$ & 2.41 $\times 10^{-4}$ & 7.86 \\
773.50 & 773.66 & 773.50 & 0.9904995 & 2.19 $\times 10^{-4}$ & 5.84 $\times 10^{-4}$ & 2.44 $\times 10^{-4}$ & 4.90 \\
777.20 & 777.36 & 777.20 & 0.9910061 & 1.99 $\times 10^{-4}$ & 6.15 $\times 10^{-4}$ & 2.47 $\times 10^{-4}$ & 6.43 \\
\hline
\multicolumn{8}{c}{Out-of-Transit} \\
\hline
768.86 & --- & 768.79 & 1.1201531 & 8.85 $\times 10^{-5}$ & 5.62 $\times 10^{-4}$ & 1.24 $\times 10^{-9}$ & 1.00 \\
770.00 & --- & 769.93 & 1.1200131 & 5.05 $\times 10^{-5}$ & 6.05 $\times 10^{-4}$ & 6.71 $\times 10^{-11}$ & 1.00 \\
773.76 & --- & 773.69 & 1.1202807 & 8.46 $\times 10^{-5}$ & 5.54 $\times 10^{-4}$ & 4.65 $\times 10^{-9}$ & 1.00 \\
777.45 & --- & 777.38 & 1.1196202 & 8.18 $\times 10^{-5}$ & 5.78 $\times 10^{-4}$ & 8.29 $\times 10^{-5}$ & 1.45 \\
\hline

\end{tabular}

\medskip
$\lambda_{E}$ is the observed wavelength from the GTC (i.e., from the Earth), $\lambda_{P}$ is the observed wavelength in the frame of the planet, and $\lambda_{S}$ is the observed wavelength in the frame of the star. Values for $\lambda_{P}$ are not given for the out-of-transit observations, as the planet was not transiting and was therefore not technically observed.  The in-transit ratios refer to the relative flux ratio between the target and reference that has been normalized to the weighted mean OOT flux ratios (given at the bottom of the table).

\end{minipage}
\end{table*}

\subsection{Effects of Earth's Atmosphere}
\label{secearth}

We consider the effect of random atmospheric variations (e.g., clouds) during the night of the transit as well as during the April baseline observations.  As mentioned in \S\ref{secRed}, large variations in the absolute flux of both the target and reference were observed towards the beginning and the end of the transit observations, with a few large fluctuations around the middle of the observations as well.  Thus, to check if our measured in-transit flux ratios were affected by these fluctuations, we computed the weighted mean in-transit flux ratio for each bandpass after excluding outlying absolute flux measurements from our analysis.  We specifically excluded any points that were greater than 3$\sigma$ away from the mean of the flattest part of the spectrum measured for each bandpass and each star.  After excluding outlying points from both the in-transit and April baseline data, we found that the new spectrum for HD 80606b shows a very similar shape as the original spectrum, albeit with the flux ratio in the reddest bandpass differing the most from the original spectrum.  However, we still measure a significant difference between the flux ratios in the on-line and reddest bandpasses.  These results are included in Table \ref{tabratio2} and shown in Figure \ref{figspectrum} as the solid circles.  

\begin{table*}
 \centering
 \begin{minipage}{140mm}
  \caption{Time-Averaged Flux Ratios and Noise Estimates (Outlying Absolute Fluxes Excluded) \label{tabratio2}}
  \begin{tabular}{@{}cccccccc@{}}
  \hline
$\lambda_{E}$ (nm) & $\lambda_{P}$ (nm) & $\lambda_{S}$ (nm) & $< \delta F/F >$ & $\sigma_{< \delta F/F >}$ & $\sigma_{w}$ & $\sigma_{r}$ & $\beta$ \\
\hline
\hline
\multicolumn{8}{c}{In-Transit} \\
\hline
768.60 & 768.76 & 768.60 & 0.9901021 & 4.83 $\times 10^{-5}$ & 4.88 $\times 10^{-4}$ & 3.90 $\times 10^{-5}$ & 1.00 \\
769.75 & 769.91 & 769.75 & 0.9901218 & 1.76 $\times 10^{-4}$ & 6.29 $\times 10^{-4}$ & 2.03 $\times 10^{-4}$ & 6.44 \\
773.50 & 773.66 & 773.50 & 0.9904292 & 2.15 $\times 10^{-4}$ & 5.82 $\times 10^{-4}$ & 2.40 $\times 10^{-4}$ & 4.80 \\
777.20 & 777.36 & 777.20 & 0.9907548 & 2.42 $\times 10^{-4}$ & 5.89 $\times 10^{-4}$ & 2.90 $\times 10^{-4}$ & 7.88 \\
\hline
\multicolumn{8}{c}{Out-of-Transit} \\
\hline
768.86 & --- & 768.79 & 1.1202527 & 1.01 $\times 10^{-4}$ & 5.66 $\times 10^{-4}$ & 7.78 $\times 10^{-9}$ & 1.00 \\
770.00 & --- & 769.93 & 1.1201327 & 1.20 $\times 10^{-4}$ & 5.89 $\times 10^{-4}$ & 1.13 $\times 10^{-4}$ & 2.19 \\
773.76 & --- & 773.69 & 1.1203722 & 8.98 $\times 10^{-5}$ & 5.77 $\times 10^{-4}$ & 1.35 $\times 10^{-9}$ & 1.00 \\
777.45 & --- & 777.38 & 1.1196849 & 5.99 $\times 10^{-5}$ & 5.82 $\times 10^{-4}$ & 3.36 $\times 10^{-9}$ & 1.00 \\
\hline

\end{tabular}

\medskip
Same as in Table \ref{tabratio}, but the flux ratios listed here are those computed after excluding outlying absolute flux measurements from the analysis.

\end{minipage}
\end{table*}

\subsection{Limb Darkening Effects}
\label{secld}

So far our analysis has assumed that LD is the same between our different bandpasses, so LD should not affect the mean flux ratios for each bandpass differently.  However, in principle, there is also the possibility that LD coefficients vary significantly in and out of narrow spectral lines.  To investigate the possibility that our spectrum's signature is a result of our probing in and out of HD 80606's stellar spectral lines, we have computed quadratic LD coefficients for each of our bandpasses for a grid of stellar models [using PHOEBE; \citet{prsa2005}].  We then generated theoretical limb-darkened light curves for each bandpass using the standard planet transit model of \citet{mandel2002}.  We used stellar parameters and uncertainties for HD 80606 as given by \citet{winn09} to estimate a range of LD coefficients to use in our models.  We also input planetary parameters and uncertainties for HD 80606b as given by \citet{hebrard10}.  After computing light curve models for different combinations of LD coefficients and planetary parameters, we computed the mean model flux ratio over the bottom of each transit light curve (the 4 hours centered around mid-transit).  We include the resulting model spectrum in Figure \ref{figspectrum} as solid squares.  This particular spectrum was computed based on using a median set of LD coefficients, but all the model results were similar over the range of LD coefficients used.  The median linear and quadratic LD coefficients ($u_1, u_2)$ are (0.392,0.229), (0.388,0.233), (0.391,0.230) and (0.376,0236) for the 768.76, 769.91, 773.66 and 777.36 nm bandpasses.

While small differences in LD exist between the different bandpasses, the mean model flux ratios differed by only a very small amount ($< 2\times10^{-5}$) between the different bandpasses.  From this, we conclude that LD is most likely not the cause of the large variations in our observed spectrum.  However, we note that PHOEBE (as well as other LD codes) has not been calibrated in and out of narrow spectral lines.  We also note that the models show that the bottom of the light curve is in fact not flat due to LD.  However, based on our calculation of the mean model flux ratio over the limb-darkened transit bottom for each bandpass, this should not affect the magnitude of the variations we measure in our observed spectrum.  Due to LD effects, the overall normalization of the spectrum may be affected.  

\subsection{Transit Colour}
\label{secColor}

\begin{figure}
\includegraphics[width=84mm]{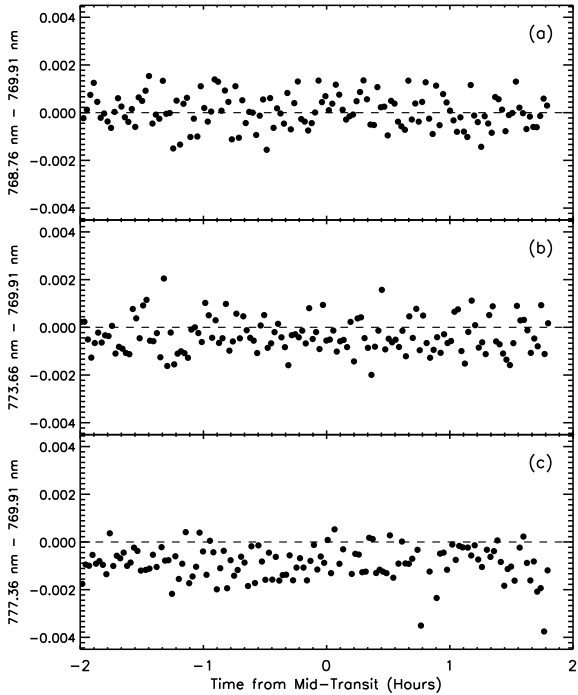} 
\caption{Colours of the normalized in-transit flux ratios.  The different panels show the colour as computed between each off-line bandpass and the on-line bandpass (after binning the on-line data to the number of points in each of the off-line bandpasses).  The dashed line in each panel illustrates where the colour equals zero.  The data has not been explicitly offset, and that there are no obvious systematics seen in any of the colours.} 
\label{figcolor}
\end{figure}

In Figure \ref{figcolor} we present the colour of the normalized in-transit flux ratios, computed by dividing each point in the off-line bandpasses by the average of each pair of on-line points around those off-line points.  We find that the colour between the bluest bandpass and the on-line bandpass is consistent with zero, with a mean value of $6.30\pm$$6.04\times10^{-5}$ (computed following the method described in \S\ref{secResults}).  The mean colour of the 773.66 nm and on-line bandpasses is $-3.57\pm$$0.63\times10^{-4}$, and the mean colour between the reddest and on-line bandpasses is $-8.99\pm$$0.62\times10^{-4}$.  

\begin{figure}
\includegraphics[width=84mm]{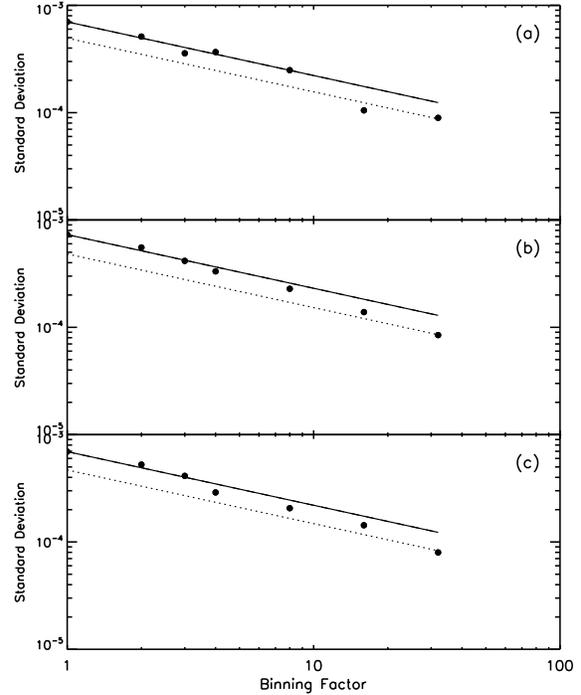} 
\caption{Standard deviation of the time-binned colour measurements from the bottom of the transit (as shown in the different panels in Figure \ref{figcolor}).  The different panels show the standard deviations for the different colours as presented in the panels in Figure \ref{figcolor}, with panels (a), (b) and (c) respectively showing the standard deviations for the 768.76 - 769.91 nm, 773.66 - 769.91 nm and 777.36 - 769.91 nm colours.  The solid line in each panel represents the trend expected for pure white Gaussian noise ($\sim N^{-1/2}$).  The dotted lines represent the trend expected for Gaussian noise when normalized to the unbinned theoretical uncertainties for these observations.  There is no obvious presence of red noise at large binning factors.} 
\label{figcolorrms}
\end{figure}

We also present the standard deviation of each colour for a number of binning factors in Figure \ref{figcolorrms}.  We find that the trend for each colour is consistent with having only white noise in each of our colours.  This is also confirmed by fitting the white and red noise explicitly for each colour.  Considering that the red noise is estimated to be less than $\sim$1$\times10^{-8}$ for each colour, white noise clearly dominates the uncertainties in the transit colour.

\begin{figure}
\includegraphics[width=84mm]{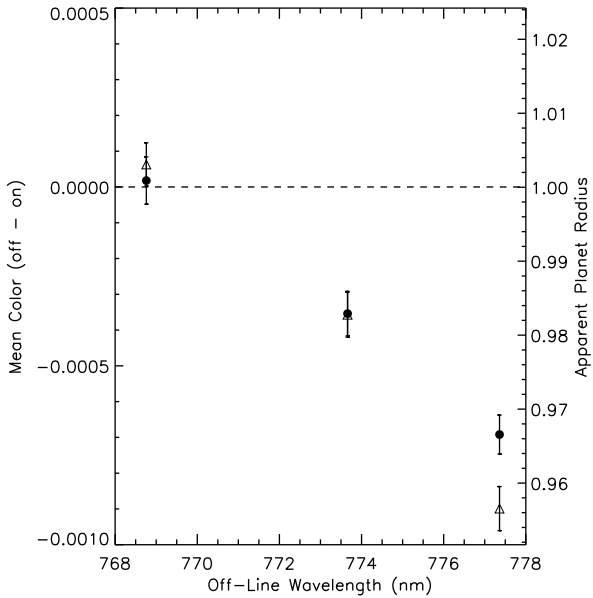} 
\caption{Mean colour of the in-transit flux ratios as computed between each off-line bandpass and the on-line bandpass.  The open triangles represent the colours as computed in \S\ref{secColor} and illustrated in Figure \ref{figcolor}.  The solid circles represent the colours computed after excluding outlying absolute flux measurements for each star from the analysis (see \S\ref{secearth}).  The errors bars represent the 1$\sigma$ uncertainties.  The dashed line illustrates where the colour equals zero.  We arbitrarily set this point equivalent to an apparent planet radius of 1 (i.e., we let the measured radius in the on-line bandpass be the baseline radius of HD 80606b).  The mean colours around the 773.66-nm bandpass are essentially equal for both sets of points, so the two data points appear as one.}   
\label{figallcolors}
\end{figure}

As explored in \S\ref{secearth}, we also compute mean colours after excluding outlying absolute flux measurements from our analysis.  After excluding those data points, we estimate the mean colours between each off-line and the on-line bandpasses to be $1.79\pm$$6.60\times10^{-5}$, $-3.54\pm$$0.62\times10^{-4}$, and $-6.92\pm$$0.54\times10^{-4}$ (from bluest to reddest).  Both these mean colours and those discussed above are plotted in Figure \ref{figallcolors}.  The colours are comparable between the two data sets, with the colour of the reddest bandpass having the only measurable difference between the two sets.  Furthermore, Figure \ref{figallcolors} illustrates that not only is there a significant change in the colour during transit, but also that the magnitude of the change is equivalent to a large change in the apparent planet radius.  At the reddest wavelengths, we clearly measure a change of over 3\% (and as much as 4.2\%, based on the flux ratios that do not exclude outlying absolute flux measurements) in the apparent radius of the planet compared to the planet's apparent radius in the on-line bandpass.

Overall, these colours agree with the magnitude and direction of the differences measured between the weighted mean in-transit flux ratios for the different bandpasses (see \S\ref{secResults}).  Furthermore, the differences between the colour of the bluest-to-on-line bandpasses and the reddest-to-on-line bandpasses has greater than $5\sigma$ significance.  Since our colour measurements match the magnitude and direction of the differences in the flux ratios as measured from our spectrum, we conclude that our measured spectrum of HD 80606b's atmosphere is real and that the differences in the flux ratio are significant.

\section{Discussion}
\label{secDiscuss}

\subsection{Interpretation of Light Curve Shape}
\label{secShape}

First, we compare our light curve (integrated over all bandpasses) to simultaneous observations from {\em Spitzer} \citep{hebrard10} and other ground-based observatories \citep{shporer10}.  In particular, \citet{hebrard10} identified a bump in the in-transit light curve that occurred within the hour $before$ their estimated time of mid-transit and pondered whether it could be due to a Moon or spot crossing.  Under the Moon hypothesis, the magnitude of the bump should be wavelength independent.  If the bump were due to a spot, then one would expect an even greater feature in the optical.  We do not find any evidence for a coincident bump [regardless of whether we adopt the ephemeris of \citet{hebrard10} or \citet{shporer10}].  Thus, the bump is unlikely to be due to either a Moon or starspot.  If anything, we find possible evidence of a bump occurring {\em after} mid-transit, but this feature was not detected by \citet{hebrard10}.  If we assume our candidate bump is not a result of instrumental systematics, and we compare our candidate bump as observed in the different wavebands, we find that the size of the putative bump is smallest in the bluest bandpass, providing further evidence against a starspot.  Furthermore, since the magnitude of the bump varies slightly for each bandpass, this provides additional evidence against the existence of a Moon.  Future high-precision, multi-wavelength observations could help provide additional constraints on the light curve shape.

\subsection{Comparison to Previous Observations}
\label{secCompare}

Next, we note that our measured in-transit flux ratios differ slightly from the flux ratio given by \citet{hebrard10}.  This is at least partly due to the different bandpasses used.  There could also be a systematic uncertainty in the overall normalization of our transit depths.  If our goal had been to measure the transit depth precisely, we would have required observations taken just before and after the transit event.  In this case, ground-based observations spanning the full transit were not feasible due to the extremely long transit duration.  Thus, we normalized our in-transit light curves by OOT observations taken on a different night.  While our observations resulted in a very high precision for differential measurements of the transit depth in each bandpass, a change in the observing conditions between nights could result in the transit depths all being affected by a common scaling factor.  

To confirm that the change in the apparent planetary radius with wavelength is based on a robust estimate of the OOT flux ratio despite using baseline observations separated by four months from the transit observations, we estimated the weighted mean in-transit flux ratios as before, but normalized them against the lower quality OOT data taken on 2010 January 15.  For reference, we include the results of the aperture photometry for this data set in Table \ref{tabjanout} and the flux ratios before and after EPD in Table \ref{tabootjan}.  We found that despite the large scatter in that OOT data, the normalized in-transit flux ratio (and therefore, the apparent radius of the planet) still changes significantly with wavelength and maintains the same shape as shown in Figure \ref{figspectrum}.  We conclude that the large change in transit depth from the 768.76 and 769.91 nm bandpasses to the 773.66 and 777.36 nm bandpasses is a robust result.  We also emphasize that the atmosphere was much more stable during the April observations than the January baseline observations, so we still rely on the April baseline observations for our primary analysis.

\begin{table}
  \caption{Absolute OOT Photometry from 2010 January 15 \label{tabjanout} }
  \begin{tabular}{@{}cccc@{}}
  \hline
$\lambda$ (nm) & HJD & $F_{target}$ & $F_{ref}$ \\
\hline
768.60 & 2455211.7489 & 1077950 & 962913 \\
\multicolumn{4}{c}{\dots} \\
\multicolumn{4}{c}{} \\
769.75 & 2455211.7487 & 1210710 & 1082178 \\
\multicolumn{4}{c}{\dots} \\
\multicolumn{4}{c}{} \\
773.50 & 2455211.7492 & 1020447 & 911377 \\
\multicolumn{4}{c}{\dots} \\
\multicolumn{4}{c}{} \\
777.20 & 2455211.7495 & 1301394 & 1161982 \\
\multicolumn{4}{c}{\dots} \\
\hline
\end{tabular}

\medskip
Columns are similar to Table \ref{tabjanin}, except the wavelengths included in the table are the wavelengths as observed from the GTC (see text for additional details).  The full table is available online, and a portion is shown here so the reader can see the formatting of the table.

\end{table}

\begin{table}
  \caption{Relative OOT Photometry from 2010 January 15 \label{tabootjan} }
  \begin{tabular}{@{}ccccc@{}}
  \hline
$\lambda$ (nm) & HJD & $F_{ratio}$ & $F_{ratio}$ & Uncertainty \\
\multicolumn{1}{c}{} & \multicolumn{1}{c}{} & \multicolumn{1}{c}{$(raw)$} & \multicolumn{1}{c}{$(corrected)$} & \multicolumn{1}{c}{} \\
\hline
768.60 & 2455211.7489 & 1.11947 & 1.12007 & 0.00128 \\
\multicolumn{5}{c}{\dots} \\
\multicolumn{5}{c}{} \\
769.75 & 2455211.7487 & 1.11877 & 1.11775 & 0.00121 \\
\multicolumn{5}{c}{\dots} \\
\multicolumn{5}{c}{} \\
773.50 & 2455211.7492 & 1.11968 & 1.11940 & 0.00131 \\
\multicolumn{5}{c}{\dots} \\
\multicolumn{5}{c}{} \\
777.20 & 2455211.7495 & 1.11998 & 1.11910 & 0.00117 \\
\multicolumn{5}{c}{\dots} \\
\hline
\end{tabular}

\medskip
The columns are similar to Table \ref{tabootapr}.  The full table is available online, and a portion is shown here so the reader can see the formatting of the table.

\end{table}

We also tested whether our results were sensitive to the aperture radius used for photometry.  We tried a variety of annuli for the apertures for both the in and out-of-transit data sets.  In all cases, we see the trend of increasing flux ratio with wavelength and found very similar results to those presented here.  The only difference occurred for the largest apertures, in which case the weighted mean in-transit flux ratio on the \ion{K}{i} feature is slightly smaller than the flux ratio at the bluest wavelength.  Even for this choice of apertures, the fluxes in the 773.66 and 777.36 nm bandpasses are not significantly different (though the error bars are slightly larger).

\subsection{Lack of a \ion{K}{I} Line Core}
\label{secInterpCore}

As illustrated in Figure \ref{figspectrum}, there is no significant difference between the observations acquired in the core of the \ion{K}{i} line and slightly to the blue.  Given the 1.2-nm FWHM, a Doppler shift of $\gtrsim$ 200 km s$^{-1}$ would be needed to shift the line core out of the on-line bandpass.  This is greater than the escape speed from HD 80606b ($\sim$121 km s$^{-1}$).  Thus, we place a 3$\sigma$ limit on the strength of the \ion{K}{i} line core of $3\times10^{-4}$ (for our 1.2-nm FWHM bandpass).

By itself, the lack of a line core is most naturally explained by a lack of \ion{K}{i} at the altitudes probed by transmission spectrophotometry.  This could occur if: 1) there is a significant bulk under-abundance of potassium, 2) the potassium has condensed into clouds and/or molecules, 3) there is a cloud or haze layer above the region capable of causing significant potassium absorption, and/or 4) the potassium has been photoionized \citep{fortney03}.  In the previous case of HD 209568b, theoretical investigations of the unexpectedly weak \ion{Na}{i} absorption showed that the observed feature depth is particularly sensitive to the extent of cloud formation \citep{fortney03}.  In the case of HD 80606b, the highly eccentric orbit results in flash heating near pericenter and extreme temperature variations over the orbital period.  At the time of transit, the star-planet separation is $\sim$0.3 AU, so the equilibrium temperature is $\sim500$ K.  Based on Spitzer observations, cooling is sufficiently rapid that the planet is expected to  have cooled between pericenter and transit \citep{laughlin09}.  Thus, both sodium and potassium are predicted to have condensed into clouds.  Thus, we conclude that the lack of a \ion{K}{i} core could easily be due to potassium having condensed into clouds before the time of transit.

\subsection{Planetary Atmosphere Models}
\label{secModel}

In an attempt to model our observations, we considered both a conventional 1-D ``cold'' atmosphere model \citep{fortney10} (solid line in Fig.\ \ref{figspectrum}) and a similar model, but with arbitrary additional heating to raise the effective temperature by 500 K (dotted line in Fig.\ \ref{figspectrum}).  Both models have been normalized to the stellar radius estimated by \citet{hebrard10}, and assume a star-planet separation of 0.3 AU (i.e., the distance between the star and planet when the planet transits).  Chemical equilibrium and a standard pressure-temperature profile for HD 80606b are assumed.  In the ``cold'' atmosphere model, the planet's (apparent) radius at 10 bar was adjusted to match the radius measured by \citet{hebrard10} at 4.5 microns.  In the ``hot'' atmosphere model, the temperatures in the upper atmosphere range from $\sim$300$-$500 K, even with the additional heating. The higher temperature increases the observed planetary radius at all wavelengths, and slightly increases the peak to trough distance of the features, but the planet's radius was not adjusted to match the radius from \citet{hebrard10}.  At these temperatures, most of the potassium is expected to have formed condensates, significantly reducing the \ion{K}{i} absorption feature.  As the inset in Figure \ref{figspectrum} illustrates, neither the ``hot'' and ``cold'' atmosphere models predict a significant feature due to \ion{K}{i} absorption.

\subsection{Change in Apparent Radius with Wavelength}

While we do not detect the \ion{K}{i} core, we find relatively large differences ($3.57\pm0.63\times10^{-4}$ and $8.99\pm0.62\times 10^{-4}$) between the colours of the on-line bandpass and the bandpasses to the red (773.66 and 777.36 nm).  Clouds and hazes would suppress both the core and wings of the absorption feature.  A similar observation for a typical hot-Jupiter could be readily interpreted as strong absorption in the wings of the potassium line due to absorption by pressure broadened potassium at lower altitudes, while potassium at higher altitudes has been photoionized \citep{fortney03}.

However, in our observations, the magnitude of the difference in absorption at the two blue and two red wavelengths appears too large for such a model.  One could expect such observations to probe the lower atmosphere over $\sim$10 scale heights ($H$), from a pressure of $\sim$100 mbar to $\sim$1 microbar.  Assuming the planet has reached a thermal equilibrium for the star-planet distance at the time of transit and a 500K upper atmosphere temperature, the scale height would be $H\sim~$20 km.  Thus, one might expect to see changes in the apparent radius of the planet on the order of $\sim~200$ km.  Our observations suggest a much larger change in the apparent radii (up to $\sim4.2$\% or $\sim 2900$ km) when comparing observations in the \ion{K}{i} line core and the reddest bandpass.  The scenario described above would suggest that these observations probed $\sim$145 scale heights in the atmosphere of HD 80606b, or pressures of less than $\sim10^{-55}$ bars, which is well into the exosphere.  Such a large number of scale heights is not realistic, implying that the absorption is originating from a part of the atmosphere much hotter than 500 K.  Fortunately, the temperature is expected to rise rapidly to thousands of Kelvin above one planetary radius \citep{yelle04}.

\subsection{Absorption by an Exosphere}
\label{secExosphere}

Based on the model of \S\ref{secModel}, we would estimate that our observations have probed $\sim$145 scale heights in the atmosphere of HD 80606b, or a pressure of less than $10^{-55}$ bars.  However, these estimates assume an atmospheric temperature of 500 K.  \citet{yelle04} finds a steep rise in the temperature from $\sim$350 to 10000 K from 1 $R_p$ to 1.1 $R_p$ for a planet at 0.1 AU from the Sun.  If we use their model as a rough guideline, and if we assume a temperature of 2000 K between 1 and $\sim$1.04 $R_p$ for HD 80606b, the 2900 km measured change in the apparent radius would imply that the observations probed $\sim$36 scale heights, or to a pressure of less than $10^{-14}$ bars.  Regardless of whether we assume a temperature of 500 K or 2000 K, the implied pressures are indicative of those that would exist in an exosphere.  

The models and opacity database of \S\ref{secModel} are not complete for the temperature and pressures of the exosphere.  The opacity database used extends to temperatures of $\sim$2600 K and $\sim$1 microbar and is not intended to describe opacity sources in an exosphere or wind \citep[e.g.,][]{vidalmadjar03, vidalmadjar08, ballester07, ehrenreich08, lecav08, lecav10, benjaffel10}.  To the best of our knowledge, an exospheric model that predicts the location and strength of absorption features arising from the exosphere does not exist.  We hope that our observations will stimulate theoretical models for the observable effects of exoplanet exospheres on transmission spectroscopy and spectrophotometry.

Given the planet's high surface gravity and any reasonable choice of planetary parameters, a $\sim4.2$\% change in the planet's apparent radius requires a very dramatic change in the pressure at which the slant optical depth reaches unity, between 770 and 777 nm.  Thus, we conclude that absorption at high altitude and temperature is the most likely explanation for the large change in the apparent planet radius.

\subsection{Possibility of Other Absorbers}
\label{secOtherAbs}

Next, we consider whether another absorber might be responsible for the observed change in apparent planet radius.  Methane can be active in this region of the spectrum.  However, methane would be unstable at the high temperatures of an exosphere or wind.  Both of the models in \S\ref{secModel} include methane at all temperatures at which it would be stable, around $<$1000 K.  In the wavelength regime that we observed, the opacity of methane is largest at 778 nm and smallest at 769 nm, so its presence would produce the opposite trend from what is shown in the data.  

The observed wavelengths were also chosen to avoid water vapor (which is also unstable at high temperatures).  We are not of aware of any other absorber which could explain the large change in apparent planet radius, and consider \ion{K}{i} the most likely absorber.  Nevertheless, we cannot rule out the possibility that HD 80606b's exosphere possesses an absorber that is something other than \ion{K}{i} on account of the incomplete opacity database.

\subsection{Absorption by a Wind}
\label{secWind}

If the $\sim4.2\%$ change in the apparent radius is due to absorption by \ion{K}{i} at high altitude, then it is not obvious why the observations on the \ion{K}{i} core (769.91 nm) are not significantly different from the observations slightly to the blue (768.76 nm).  One possibility is that the line core was shifted out of the on-line bandpass.  Given the $\sim1.2$-nm FWHM, this would require a Doppler shift of $\gtrsim200$ km s$^{-1}$.  A blue shift of $225$-km would place the core halfway between the 768.76 and 769.91 nm bandpasses.  A somewhat smaller Doppler shift plus Doppler broadening might also reduce the signal strength.  In any case, the velocities required would be greater than the escape speed from HD 80606b ($\sim$121 km s$^{-1}$).  
  
While a velocity exceeding the escape speed is somewhat concerning, it is not out of the question for a wind being driven from the exosphere.  In fact, similar observations of other planets also appear to find an unexpectedly large Doppler shift.  Specifically, a large blue shift has been found in all cases.  E.g., \citet{redfield08} found an unexpected blueshift of the core of the \ion{Na}{i} absorption for HD 189733b.  \citet{snellen10} detected a 2 km s$^{-1}$ blueshift in the upper atmosphere of HD 209458b with observations of CO.  Additionally, \citet{holmstrom08} reported Lyman-$\alpha$ absorption around HD 209458b at wavelength offsets corresponding to  velocities of several 100 km s$^{-1}$, but there was no information about the Lyman-$\alpha$ core as it is not observable due to Earth's geo-corona.  Much like our observations of HD 80606b, there is considerable uncertainty regarding the origin of the absorption and Doppler shift for HD 209458b \citep{lecav08,lecav10,benjaffel10}.  Proposed mechanisms include radiation pressure and interaction with a stellar wind \citep[e.g.,][]{tian05, garciamunoz07, murrayclay09, ekenback10}, and in particular we note that models of HD 209458b's atmosphere match observations better if it is assumed that the line core is obscured.  Our observations could be explained if a similar mechanism operates on HD 80606b and heavy elements (i.e., potassium) are mixed into the wind.
 
In the case of HD 80606b, the dynamics of the exosphere and any planetary wind is almost certainly quite complex.  The planet has the  largest semi-major axis of any confirmed transiting planet ($0.455$ AU),  but it follows such a highly eccentric orbit ($e=0.93$) that the star-planet separation of HD 80606b at periastron is $\sim$2/3 that of HD 209458b.  Thus, HD 80606b experiences strong and rapid heating of the atmosphere near pericenter.  The large and rapid changes in the incident stellar flux and temperature, as well as the stellar wind flux could lead to episodic mass loss following each pericenter passage \citep{laughlin09}.  Based on the observed X-ray flux \citep{kashyap08} and mass loss correlation \citep{wood05}, HD 80606 could have a mass loss rate as much as $\sim$100$\times$ stronger than HD 209458, providing a much stronger stellar wind to drive a wind from HD 80606b.  The rapid contraction and expansion of the Roche lobe around each pericenter could further complicate the dynamics of the exosphere and planetary wind.

\subsection{Potential Systematics}
\label{systematics}

\subsubsection{Excluding Telluric Absorption}
\label{telluric}

The usual suspect in ground-based observations is variability in the telluric absorption.  At our observed wavelengths there is very little absorption.  The only two species that contribute any appreciable absorption are water and oxygen.  In particular, there is a lack of absorption from carbon dioxide or methane in our observed bandpasses.  Oxygen is generally well mixed in the atmosphere.  Thus, we expect any variability due to oxygen has been removed in our data reduction procedure, which normalizes each observation of HD 80606 by the flux of HD 80607 taken at the same time and using the same bandpass.  Thus, we rule oxygen absorption out as a potential systematic.

Since water can be very anisotropically distributed in the atmosphere, one could worry that the 20 arcsec separation between HD 80606 and HD 80607 might allow for variations in the water absorption that are not removed by calibration.  However, the two bandpasses to the red of \ion{K}{i} were specifically chosen to be at wavelengths that avoid water absorption.  Thus, even in the scenario that the on-line and blue bandpasses were contaminated by water absorption, we still measure a $\sim2.7$\% change in the apparent planet radius between the two reddest bandpasses (both of which should be substantially free of telluric absorption).  From this, we conclude that our primary result of measuring a large change in the apparent radius with wavelength is not the result of variable telluric water absorption.

However, in an effort to confidently rule variable telluric absorption out as a potential source for systematics, we construct an alternative model for the spectrum based on changing the level of water vapor absorption.  Specifically, we integrated our bandpasses over high-resolution model transmission spectra for telluric water vapor and oxygen.  Using the TERRASPEC code (Bender et al. 2011, in prep), we computed model transmission spectra for two different airmasses (representing the mean airmass over the transit bottom as observed in January and the mean airmass during the baseline data observed in April) and three different water vapor levels (1, 5 and 10 mm).  We then integrated our bandpasses over each spectra and computed the relative transmission for the different bandpasses for every possible combination of water vapor towards HD 80606 and HD 80607.  We integrated over the appropriate bandpasses for each set of observations, as the bandpasses used in April were centered at slightly different wavelengths than for the January observations.  Our goal was to determine if the transmission spectrum would have a similar signature as our observed spectrum if there was a difference in the water vapor towards HD 80606 and HD 80607 during either or both of the January and April observations.  For example, we took the integrated transmission for a water vapor level of 10 mm (towards HD 80606) divided by the integrated transmission for a water vapor level of 1 mm (towards HD 80607) based on the mean airmass in January.  Then, we divided that result by a similar ratio based on the spectra for the April observations.  We computed this ratio for all combinations of water vapor and compared the results. Realistically, the water vapor was most likely below 6 mm for both the January and April observations [based on \citet{garcia2010}], but we approach this issue with much caution and therefore discuss the results for the 10 mm water vapor level as well.

From our results, we can make several arguments against variable water vapor absorption and/or the different wavelengths observed in January and April being the cause of our spectrum's signature.  First, since our measurements are multiply differential (comparing the target to the reference in-transit to the target to the reference out-of-transit), we minimize any such effects from our measurements.  Second, even if the water vapor column towards HD 80606 and HD 80607 differed by an average of ~10 mm on one of the nights, it would result in a difference of only 0.0058\% (if the January water vapor differed by 10 mm) or 0.034\% (if the April water vapor differed by 10 mm) between the reddest and on-line bandpasses.  This difference in transmission based on the wavelengths observed in April is less than half of the actual measured difference between the flux ratios in these two bandpasses.  Further, the separation between HD 80606 and HD 80607 is only 20 arcsec, so it is extremely unlikely that the time-averaged water column towards the two stars would differ by 10 mm.  Further, an untenably large water column, inconsistent with the observational conditions, would be required to explain the observed difference of $\sim$8$\times10^{-4}$ between the on-line and reddest bandpasses.  Third, even if the average water column towards the two stars did differ by that much on one of the nights, the resulting colours differ from what we observe.  I.e., the hypothesis that our measurements are primarily due to atmospheric variability would predict the two bluest bandpasses to be comparable in some cases and differ largely in others, while the two reddest bandpasses are comparable in all cases.  While we do observe the flux ratios in the two bluest bandpasses to be comparable, we see a significant difference between the flux ratios for the two reddest bandpasses.  Further, the magnitude of the differences between the bluest and reddest bandpasses is observed to be much larger than what the difference would be if they were caused by variable atmospheric absorption. 

Thus, we estimate that for the four different bandpasses, the effect of variable atmospheric absorption would be less than (0.0024\%, 0.0015\%, 0.0075\%, 0.0073\%) $\times$ [ $<$ mm of H$_2$O towards HD 80606 $ >$ - $<$ mm of H$_2$O towards HD 80607 $ >$ ]/[10 mm of H$_2$O] at the wavelengths and airmass observed at in January, or less than (0.004\%, 0.027\%, 0.0041\%, 0.0068\%) multiplied by the same ratio given above at the wavelengths and airmass observed at in April.  However, assuming that the difference in transmission is neglible between the target and reference for both the January and April observations, then the fact that the stars were observed at different wavelengths and airmasses on those nights should be irrelevant.

Finally, we note that spectrophotometry using a narrow-band tunable filter is much less prone to systematics than spectroscopic observations.  The lack of a slit, the simultaneous use of a very good reference star, rapid switching between bandpasses, and the multiply differential nature of our measurement should all minimize the effects of telluric absorption.  While the OH lines are variable, the sky subtraction in our data reduction process removes the emission to a high degree of precision.  Finally, we see no evidence, in our atmospheric transmission models, of absorbers that could account for the signal detected.

To first order, the effects of atmospheric extinction are corrected by measuring flux ratios relative to HD 80607.  We expect negligible second order differential extinction, since the target and reference stars are of the same spectral type and separated by only 20 arcsec.  Since the magnitude of this effect scales as the square of the filter bandpass, our use of such a narrow bandpass further minimizes second order differential extinction, allowing this technique to be applied to other targets with reference stars that differ in temperature.

We do not consider differential extinction to be a viable explanation for the effect seen in Figure \ref{figspectrum}.  Nevertheless, we performed an additional check, in which we do not perform relative photometry between the target and reference.  We compare the ratio of the absolute flux of the reference star in the reddest bandpass and the bandpass centered on \ion{K}{i} as measured on the night of the transit to the same ratio as measured on the night the OOT observations were taken (2010 April 4).  We estimate a ratio of 0.98498 $\pm$ 0.00093, equivalent to a colour deviation of $\sim$1.5\% between the two nights.  This provides an upper bound on the effects of atmospheric variability, including differential extinction.  The accuracy of our primary analysis should be considerably higher thanks to the use of relative photometry to correct for atmospheric variability.

\subsubsection{Excluding Instrumental Effects}

With tunable filter imaging, the photons for each observed wavelength lands on the same pixels, eliminating concerns about spatial variations in the flat fielding.  However, the normalization of the flux measurements is affected by the wavelength-dependence of the pixel sensitivity.  To minimize this effect, we took dome flat fields for each observed wavelength and corrected the science frames taken at each wavelength with their respective flat fields.

Furthermore, we guard against possible non-uniformity in the shutter motion, which could result in the systematic effect of producing slightly different exposure times for the target and reference star, depending on where they are located on the CCD chip.  This systematic effect is more noticeable for shorter exposure times, so we guard against it by following an observing sequence that repeats after seven exposures, so that the subsequent set of exposures occurs with the shutter motion in the opposite direction.

Depending on the orientation of the tunable filter, the observed wavelength can drift due to the rotation of the instrument during the observations.  For our observations, the tunable filter was tuned before observations, in the middle of the transit and at the end of the observations.  No drifts larger than 0.1-nm occurred.  

Finally, we conducted a thorough investigation into the possibility of saturation and/or non-linearity as a source of systematic effects.  A majority of the peak counts during our observations were well below the saturation threshold ($\sim$65,000 ADUs), and for standard observing modes linearity is guaranteed up to $\sim$65,000 ADUs, so non-linearity should not be an issue.  However, as we use a non-standard observing mode on OSIRIS in order to read out the CCDs at the highest rate possible (and thereby greatly reduce dead time), it is worthwhile to investigate whether non-linearity is an issue.  Therefore, we discuss here several checks for non-linearity, where we arbitrarily assume that 45,000 counts ($\sim$30,821 ADUs, based on the gain of 1.46 $e^{-}$ per ADU) is the level at which non-linearity might begin.

First, we checked if the average number of counts from the flat-fields taken for each bandpass had a linear dependence with exposure time, as we had taken flat-fields at several different exposure times.  We fit a line to all measurements of the mean flat-field counts (for 5 different exposure times), and then we fit another line to the data but excluded measurements that were near or above 45,000 counts.  To see if including measurements at higher counts resulted in non-linearity, we compared the slopes and y-intercepts of the two best-fit lines.  After comparing the best-fit solutions between the different bandpasses and for the different series of flats taken in January and April, we find that it is not obvious that any one set of flats displays significant non-linearity compared to the others.

Second, we investigated fitting a quadratic function to the flat-field counts for both the January and April flat-fields.  After comparing the best-fit coefficients for the different bandpasses, we found that at least one set of coefficients deviated significantly from the coefficients for the other bandpasses.  While there might be an obvious outlier in terms of one bandpass that might be affected by non-linearity for each set of flats, the supposed outlier is different for the January and April flats.  We again conclude that it is not obvious which, if any, of our bandpasses is displaying significant non-linearity.

Third, we investigated the possibility of non-linearity by computing the colour (between each off-line bandpass and the on-line bandpass) and seeing how it varied with respect to the average on-line flux per pixel (estimated by dividing the total absolute on-line target flux by the target's FWHM squared).  We computed the median colour for exposures where the average flux per pixel was below 45,000 and for exposures where the average counts were above 45,000.  We then estimated the difference in the median colour for those two sets of exposures.  We found that in the near-red bandpass (773.66-nm), non-linearity most likely does not play a role, as the median colours below and above the 45,000 count level differ by an insignificant amount.  However, in the bluest and reddest colours, we do see a slight correlation, with the median colours differing by comparable amounts.  This is not what we might expect to see if non-linearity were causing systematic effects in our observed spectrum, since we do not observe a comparable colour difference in the in-transit flux ratio between the blue-on-line and reddest-on-line colours.  

We also computed the colour deviations for the April baseline data, and we found that the colours below and above the 45,000 count mark are slightly larger than the in-transit colour deviations (but these were computed using a combined data set for two different exposure times, which could affect these estimates).  Regardless, we still find that the smallest difference in the colours is in the near-red bandpass, and the differences are comparable for the bluest and reddest bandpasses, even though in both the observed in-transit and out-of-transit flux ratios we see the smallest difference in the flux ratio between the bluest bandpasses and the largest between the on-line and reddest bandpass.

In summary, we conducted several checks for non-linearity.  We conclude that any effects of non-linearity are either too insignificant to affect our photometry or they are not correlated with the data.

\subsubsection{Possible Non-Planetary Astrophysical Effects}
\label{spots}

Lastly, we consider potential astrophysical systematics such as stellar variability.  Observations in all four bandpasses were obtained during the {\em same} transit.  If observations using different bandpasses had been made during different transits, then the interpretation would be ambiguous, as stellar variability (e.g., spots that the planet does not necessarily pass over) could result in apparent changes in the in-transit flux ratio.  For the large change in apparent radius to be due to stellar variability, there would need to be a $\sim4.2$\% change in the colour of either the target or reference star.  Such large variability over a small range of wavelengths is {\em a priori} unlikely for solar-like stars \citep{hebrard10}.  However, as suggested by the referee, we have estimated how spotted HD 80606 would have to be to cause a difference of $\sim$8$\times10^{-4}$ in the flux ratios in the on-line and reddest bandpass.  

We computed the blackbody flux for HD 80606 ($T_{eff}$ $\sim$ 5572 K) then integrated the flux over each bandpass to estimate the total flux observed in each bandpass.  We then completed similar calculations for a spot assuming a temperature 1000 K cooler than HD 80606 and a spot radius equal to the planet's radius.  After computing the ratio of the integrated spot flux to the integrated star flux for some $N$ spots, we found that about 26 spots with the above properties would have to exist on the surface of HD 80606 during the transit observations in order to produce a difference in the on-line and reddest flux ratios of about 8$\times10^{-4}$.  That is equivalent to having $\sim$26\% of HD 80606's surface covered with spots.  Even if the systematic trends we see in our transit light curves are due to spots coming in and out of view on the surface of the star, the percent of the stellar surface covered by spots is unlikely to be as much as 26\%.  Furthermore, if the star was this spotted, we should also see a difference in the flux ratio between the two bluest bandpasses of over 1$\times10^{-4}$, yet the difference we observe is less than $\sim$6$\times10^{-5}$.

We conclude that it is possible for spots to account for some of the variations we measure, but that HD 80606 is {\em very} unlikely to be spotted enough to cause the {\em magnitude} of variations we measure.  In fact, \citet{wright04} measured values of $S_{HK} = 0.149$ and $log$ $R^{'}_{HK}=-5.09$ for HD 80606, which indicate that the star is quite inactive.  Also, \citet{hebrard10} monitored HD 80606 and determined it is not an active star.  Specifically, they estimate that the star is photometrically stable at the level of a few mmag in the optical range on the timescale of several weeks.  Despite these statements, they still attribute the bump in their light curve to a spot on the stellar surface.  Considering the precision of our observations (much better than 1 mmag), it is possible that we observed flux variations that they did not have the precision to.  

As a final comment, we note that spots could affect the normalization of the overall spectrum, as the spectrum could need to be scaled downward (i.e., decrease the flux ratios or increase the transit depths) to account for the effect of spots.  However, the shape of the spectrum would remain the same, unless over $\sim$26\% of the star's surface was covered with spots during transit.  As noted by the referee, a large, long-lived polar spot could exist on HD 80606, which would not induce large photometric variations but could still affect our photometry.  Or, both HD 80606 and HD 80607 or HD 80607 alone could be spotted and cause the observed variations.  Due to the possible variable nature of HD 80606 (or HD 80607), we encourage future out-of-transit observations of HD 80606 and HD 80607 to determine if such variability is common.

\section{Conclusion}
\label{secConclusion}

In summary, our observations do not match existing models, due to two basic observations.  We find a large change in apparent planet radius with wavelength, but do not observe a significant difference where the \ion{K}{i} line core would be expected.  Our observations place a strong limit on the strength of the line core (unless it has been Doppler shifted by $\gtrsim100$ km s$^{-1}$), yet imply large variations in radius over wavelengths usually dominated by \ion{K}{i} absorption.  In the absence of other viable absorbers, absorption by \ion{K}{i} remains the most viable explanation.  The atmospheric scale height of HD 80606b at transit ($\sim$20 km) is significantly smaller than that of HD 209458b and HD 189733b, yet the variation in radius is larger than that of HD 209458b \citep{sing08}.  One possible model is absorption by potassium that is part of a high speed wind coming off the exosphere.   While high speed winds have been observed for other exoplanets, the mechanism for powering such winds is unclear.  We encourage further theoretical investigations to improve models for transmission spectroscopy of exoplanet exospheres in general and the specific challenge of HD 80606b.

Finally, we have investigated several potential sources of systematic effects.  There is no simple or obvious source causing the systematics in our data.  Further, any systematics introduced by the sources we have investigated here produce neither the same signature as our observed spectrum, nor the same magnitude of difference as that of our measured flux ratios.  While we are confident that none of these possible sources of systematic effects cause the shape of our observed spectrum, we still allow the possibility that one or some combination of these systematics may affect our measurements and/or the overall normalization of the observed spectrum.  We also acknowledge that the target was observed at a slightly different set of wavelengths in January as compared to April.  While the difference in wavelengths is small ($\sim$0.1-nm), there is still the possibility that this could result in small differences in either the telluric absorption or stellar spectra, which in turn could cause the observed spectrum that we have attributed to absorption from the atmosphere of HD 80606b.  As a final note, we highly encourage follow-up transit observations of HD 80606b to confirm the signal measured here.  We note that the next partial transit observable from La Palma occurs on 2012 March 3, during which observations pre-transit through the complete first half of the transit will be possible.

\section{Future Prospects}
\label{secFuture}
 
Future transit observations at wavelengths around \ion{K}{i} in HD 80606b are possible, but require considerable patience due to the long orbital period (111 days).  Observations of the transit depth around other absorbing species could test the exosphere and wind models.  Similar observations of other planets would enable a comparison of \ion{K}{i} strength in both the wing and core as a function of star and planet properties.  We note that shortly before submission, we became aware of independent, but similar, observations of another exoplanet \citep{sing10}.  Both these and future observations of additional exoplanets will enable comparisons of the atmospheric composition and structure, as well as studies of potential correlations with other planet or host star properties.  Such observations would also help improve the interpretation of the existing HD 80606b observations.

Currently, only the OSIRIS red tunable filter (651-934.5 nm) is available at the GTC.  Once the blue tunable filter is available for scientific observations, it will be possible to observe additional atmospheric features, including the \ion{Na}{i} feature previously detected for HD 209458b and HD 189733b.  The large aperture of the GTC makes it practical to perform similar observations of several fainter host stars.  Thus, we look forward to future observations of a large sample of transiting planets.  The striking diversity of exoplanets suggests that it will be fruitful to compare \ion{Na}{i} and \ion{K}{i} observations to identify trends with stellar and planet properties.

Despite the complex interpretation of these observations, the very-high-precision obtained with the OSIRIS narrow-band tunable filter imager opens up new avenues of research for large ground-based observatories.  Indeed, the measured precision exceeds that of Spitzer \citep{hebrard10} and even the Hubble Space Telescope observations for the given bandpass \citep{pont08}.  Thus, ground-based observations can now characterize the atmospheres of giant planets using spectrophotometry.  The photometric precision is also sufficient to measure emitted and/or scattered light during occultation at multiple near-infrared wavelengths that could improve constraints on atmosphere models of short-period giant planets.  By providing high-precision photometry at multiple wavelengths during a single transit, the technique could also contribute to the confirmation of transiting planet candidates, such as those identified by Kepler \citep{borucki10}.  The technique could also improve measurements of the impact parameter and thus orbital  \citep{colon2009}.  This would be particularly valuable for systems with multiple transiting planets \citep{steffen10} for which the orbital evolution depends on the relative inclination of the orbits \citep{ragoz10}.

Since all Neptune and super-Earth-sized planets will have relatively low surface gravities, they can make good targets for transmission spectroscopy.  Despite a smaller transit depth than giant planets, the potentially large atmospheric scale height can lead to a substantial signal in transmission \citep{charb09}, particularly for Neptune-sized planets orbiting sub-solar-mass stars and/or super-Earth-sized planets orbiting low-mass stars.  Previously, it has generally been assumed that the Earth's atmosphere will prevent ground-based facilities from achieving the high-precisions necessary to measure biomarkers on super-Earth-sized planets and that the James Webb Space Telescope will provide the first opportunity to characterize atmospheres of super-Earths \citep{deming09}.  If the challenges of Earth's atmosphere could be overcome, then ground-based observatories have several advantages (e.g., much larger collecting area, more modern and sophisticated instrumentation, ability to adjust and upgrade instruments).  These observations demonstrate that ground-based narrow-band photometry on large telescopes can deliver the precision necessary to characterize super-Earth-size planets around bright, nearby, small stars.  We encourage astronomers to consider a future generation of instruments specifically designed for high-precision transit observations, which may allow the characterization of super-Earth-sized planets in upcoming large ground-based observatories [e.g., Giant Magellan Telescope (GMT), Thirty Meter Telescope (TMT), and Extremely Large Telescope (ELT)].

\section*{Acknowledgments}

We gratefully acknowledge the observing staff at the GTC and give a special thanks to Ren\'e Rutten, Antonio Cabrera Lavers, Jos\'e Miguel Gonz\'alez, Jordi Cepa Nogu\'e and Daniel Reverte for helping us plan and conduct these observations successfully.  We are very grateful to Andrej Pr{\v s}a for computing limb darkening coefficients for our unique bandpasses.  We especially appreciate feedback from Dave Charbonneau, Scott Gaudi, Matthew Holman, Heather Knutson, Ruth Murray-Clay, Sara Seager, Brandon Tingley and Josh Winn.  We also thank the anonymous referee for helping us greatly improve this manuscript.  K.D.C.\ would like to acknowledge support by the National Science Foundation Graduate Research Fellowship under Grant No. DGE-0802270.  S.R. would like to acknowledge support by the National Science Foundation under Grant No. AST-0903573.  HJD acknowledges support by grant ESP2007-65480-C02-02 of the Spanish Ministerio de Ciencia e Innovacion.  The Center for Exoplanets and Habitable Worlds is supported by the Pennsylvania State University, the Eberly College of Science, and the Pennsylvania Space Grant Consortium.  This work is based on observations made with the Gran Telescopio Canarias (GTC), installed in the Spanish Observatorio del Roque de los Muchachos of the Instituto de Astrof\'isica de Canarias, on the island of La Palma.  The GTC is a joint initiative of Spain (led by the Instituto de Astrof\'isica de Canarias), the University of Florida and Mexico, including the Instituto de Astronom\'ia de la Universidad Nacional Aut\'onoma de M\'exico (IA-UNAM) and Instituto Nacional de Astrof\'isica, \'Optica y Electr\'onica (INAOE).

\label{lastpage}

\end{document}